\documentclass[epj]{svjourmod}

\usepackage{amsmath,amssymb,amsfonts}
\usepackage{graphicx}
\usepackage{cuted}

\def\Xint#1{\mathchoice
   {\XXint\displaystyle\textstyle{#1}}%
   {\XXint\textstyle\scriptstyle{#1}}%
   {\XXint\scriptstyle\scriptscriptstyle{#1}}%
   {\XXint\scriptscriptstyle\scriptscriptstyle{#1}}%
   \!\int}
\def\XXint#1#2#3{{\setbox0=\hbox{$#1{#2#3}{\int}$}
     \vcenter{\hbox{$#2#3$}}\kern-0.5\wd0}}
\newcommand{\dashint}[1]{\Xint{\hspace{#1}-}}

\allowdisplaybreaks[1]

\newcommand{\beq}{\begin{equation}}
\newcommand{\eeq}{\end{equation}}
\newcommand{\beqa}{\begin{eqnarray}}
\newcommand{\eeqa}{\end{eqnarray}}
\newcommand{\bal}{\begin{align}}

\newcommand{\mpi}{M_{\pi}}

\newcommand{\diff}{\text{d}}
\newcommand{\eps}{\epsilon}
\newcommand{\Order}{\mathcal{O}}
\newcommand{\tils}{\tilde{s}}
\newcommand{\tilt}{\tilde{t}}
\newcommand{\tilu}{\tilde{u}}
\newcommand{\mv}{m_{\rm V}}
\newcommand{\ma}{m_{\rm A}}
\newcommand{\mt}{m_{\rm T}}
\newcommand{\mta}{m_{\rm T_A}}
\newcommand{\tm}{t_{\rm m}}
\newcommand{\sm}{s_{\rm m}}
\newcommand{\qq}{\mathbf{q}}
\newcommand{\qt}{\mathbf{q}_t}
\newcommand{\pt}{\mathbf{p}_t}
\newcommand{\Lagr}{\mathcal{L}}
\renewcommand{\Im}{\text{Im}\,}
\renewcommand{\Re}{\text{Re}\,}

\begin{document}

\title{Roy--Steiner equations for $\boldsymbol{\gamma\gamma\to\pi\pi}$}
\titlerunning{Roy--Steiner equations for $\gamma\gamma\to\pi\pi$}

\author{Martin Hoferichter\inst{1,2} \and Daniel R.~Phillips\inst{2} \and Carlos Schat\inst{2,3}}
\authorrunning{M.~Hoferichter \and D.~R.~Phillips \and C.~Schat}

\institute{
   Helmholtz-Institut f\"ur Strahlen- und Kernphysik (Theorie) and Bethe Center for Theoretical Physics, Nussallee 14--16, Universit\"at Bonn, D--53115~Bonn, Germany \and Institute of Nuclear and Particle Physics and Department of Physics and Astronomy, Ohio University, Athens, OH 45701, USA
    \and Instituto de F\'{\i}sica de Buenos Aires, CONICET - Departamento de F\'{\i}sica, FCEyN, Universidad de Buenos Aires, Ciudad Universitaria, Pab.~1, (1428) Buenos Aires, Argentina}

\date{%Received: date / Revised version: date
}

\abstract{Starting from hyperbolic dispersion relations, we derive a system of Roy--Steiner equations for pion Compton scattering that respects analyticity, unitarity, gauge invariance, and crossing symmetry. It thus maintains all symmetries of the underlying quantum field theory. To suppress the dependence of observables on high-energy input, we also consider once- and twice-subtracted versions of the equations, and identify the subtraction constants with dipole and quadrupole pion polarizabilities. Based on the assumption of Mandelstam analyticity, we determine the kinematic range in which the equations are valid.  As an application, we consider the resolution of the $\gamma\gamma\to\pi\pi$ partial waves by a Muskhelishvili--Omn\`es representation with finite matching point. We find a sum rule for the isospin-two $S$-wave, which, together with chiral constraints, produces an improved prediction for the charged-pion quadrupole polarizability $(\alpha_2-\beta_2)^{\pi^\pm}=(15.3\pm 3.7)\cdot 10^{-4}~{\rm fm}^5$. We investigate the prediction of our dispersion relations for the two-photon coupling of the $\sigma$-resonance $\Gamma_{\sigma \gamma \gamma}$.
The twice-subtracted version predicts a correlation between this width and the isospin-zero pion polarizabilities, which is largely independent of the high-energy input used in the equations. Using this correlation, the chiral perturbation theory results for pion polarizabilities, and our new sum rule, we find $\Gamma_{\sigma\gamma\gamma}=(1.7\pm 0.4)\,{\rm keV}$. }


\maketitle

\section{Introduction}
\label{sec:introduction}

The reaction $\gamma \gamma \to \pi \pi$ is of particular interest in the realm of non-perturbative QCD, as it probes strong-interaction dynamics in the $0^{++}$ channel, which has the same quantum numbers as the QCD vacuum. However, a theoretical
understanding of the dynamics in this channel has long proven elusive. The pions produced in the fusion of two photons are strongly interacting, such that a description of the reaction $\pi \pi\to \pi \pi$ is a prerequisite for understanding $\gamma \gamma\to\pi \pi$.
 A significant step forward in this direction was made in \cite{CCL}, where it was shown how combining constraints from the analyticity, unitarity, and crossing symmetry of relativistic
quantum field theory (the Roy equations \cite{Roy}) with the chiral symmetry of QCD (using Chiral Perturbation Theory, ChPT \cite{Weinberg79,GL84}) produced detailed information on pion--pion amplitudes in this
channel. As a result a very precise prediction of the pole mass $m_\sigma=M_\sigma-i \Gamma_\sigma / 2$ of the $\sigma$ resonance---the lowest-lying resonance in QCD---was obtained
\beq
M_\sigma=441^{+16}_{-8}\,{\rm MeV},\qquad \Gamma_\sigma=544^{+18}_{-25}\,{\rm MeV}\label{sigma}.
\eeq
\begin{sloppypar}
This resonance influences the cross sections in $\gamma \gamma \rightarrow \pi \pi$, 
which therefore provides an alternative to meson--meson scattering reactions for its excitation. 
Experimentally, $\gamma\gamma \to\pi\pi$ is accessible in $e^+e^-$ colliders via the reaction $e^+e^-\to e^+e^-\pi\pi$, where both the incoming electron and positron radiate one photon \cite{CB90,Mark2,CELLO,Belle07,Belle08,Belle09}. However, due to its very large width, the $\sigma$ is only manifest as a broad bump in the $\gamma\gamma \to\pi^0\pi^0$ cross section, which makes it difficult to extract information on the resonance from these data. Therefore, an improved theoretical understanding of this important reaction that respects all available constraints is strongly called for.
\end{sloppypar}

Moreover, the two-photon coupling of the $\sigma$ influences Compton
scattering from the proton, via the possibility to exchange degrees of
freedom corresponding to the resonance  between the incoming photon and
the target proton \cite{Lv97,Dr02,Bernabeu_08,Schumacher10}. Developing a theory of 
$\gamma \gamma \rightarrow \pi \pi$ that
includes the $0^{++}$ resonance, and delineates its influence on
cross-section data in this channel, is therefore an important step. 

Besides its relation to nucleon polarizabilities, there has been particular interest in the two-photon width of the $\sigma$ as inferences are then made concerning the nature of this state. Apart from an interpretation as a $q\bar q$ state, possibilities such as a tetraquark state, a meson--meson molecule, or a gluonic resonance have been suggested in the literature. The coupling to two photons has been used to discriminate between different scenarios (see \cite{Pennington06,Mennessier:2008} and references therein). Extraction of the $\sigma$'s two-photon width $\Gamma_{\sigma \gamma \gamma}$ from data on $\gamma \gamma \to \pi \pi$ thus becomes an important piece of this puzzle. 
A recent $K$-matrix approach to the extraction of $\sigma$ widths from $\gamma\gamma \to\pi\pi$ data can be found in \cite{Mennessier:2008,Mennessier:2010}. Alternatively, model-independent $\gamma\gamma \to\pi\pi$ dispersion relations have been obtained and solved by means of Omn\`es techniques \cite{Muskhelishvili,Omnes} in descriptions of this reaction \cite{Gourdin_60,Babelon76,MP_88,Donoghue93,Drechsel:1999rf,FK_06,Pennington06,Pennington08,ORS_08,OR_08,Bernabeu_08,Mao09,GM_10}. The most sophisticated such treatment was in \cite{GM_10}, where, motivated by the fact that most of the Belle data lie above $1\,{\rm GeV}$, a Muskhe\-lishvili--Omn\`es representation was constructed that dynamically includes the $K \bar K$ channel, in order to obtain a description of $\gamma\gamma \to\pi\pi$ valid up to $1.3\,{\rm GeV}$. 

In this work, we consider a more general approach, namely a complete system of Roy--Steiner equations for $\gamma\gamma\to\pi\pi$ and the crossed reaction $\gamma\pi\to\gamma\pi$ that, in analogy to the $\pi\pi$ Roy equations, fully respects analyticity, unitarity, and crossing symmetry of the scattering amplitude. We find that---at a similar level of rigor at which the Roy equations for $\pi\pi$ scattering hold---our $\gamma\gamma\to\pi\pi$ equations are valid up to $1\,{\rm GeV}$. (The domain of validity can be extended further under certain additional assumptions.)
We compare our equations for the $\gamma\gamma\to\pi\pi$ partial waves to existing dispersive descriptions of this process, and find, in particular, that a numerically important coupling between $S$- and $D$-waves has been previously neglected. Furthermore, our equations lead to sum rules for the isospin-two partial waves, which we use to improve the ChPT prediction of the charged-pion quadrupole polarizability. 

As an application of these results, we study the constraints of our equations on the two-photon coupling of the $\sigma$. The subtraction constants necessary to ensure sufficiently fast convergence of the dispersion integrals can be directly related to the pion polarizabilities, which therefore play a similar role to that of the $\pi\pi$ scattering lengths in the case of $\pi\pi$ Roy equations. As the tension between various experimental determinations of the dipole polarizability of the charged pion (based on Primakov measurements \cite{Antipov82} or radiative pion production \cite{Lebedev,MAMI}) and ChPT predictions \cite{Bijnens88,DH88,BGS94,Burgi96,GIS05,GIS06} is far from being resolved, we provide the two-photon width of the $\sigma$ as a function of the pertinent polarizabilities. We are confident that the ongoing measurements at COMPASS \cite{COMPASS_1,COMPASS_2}, in combination with ChPT predictions, will clarify the situation in the near future.

The paper is organized as follows: having specified our conventions in Sect.~\ref{sec:formalism}, we present the detailed derivation of the Roy--Steiner system in Sect.~\ref{sec:roy_steiner}. The domain of validity of these equations is studied in Sect.~\ref{sec:dom_val}. We then concentrate on the equations for the $\gamma\gamma\to\pi\pi$ partial waves, whose solution in terms of Omn\`es functions is discussed in Sect.~\ref{sec:omnes}. Establishing the connection to the two-photon width of the $\sigma$ in Sect.~\ref{sec:sigma}, we then discuss the input we use and present our numerical results in Sects.~\ref{sec:input} and \ref{sec:results}. Various details of the calculation are relegated to the appendices.

\section{Formalism}
\label{sec:formalism}

\subsection{Kinematics}
\label{sec:kinematics}

We first consider the Compton-scattering process
\beq
\gamma(q_1,\lambda_1)\pi^a(p_1)\rightarrow \gamma(q_2,\lambda_2)\pi^b(p_2)
\eeq
with momenta as indicated, photon helicities $\lambda_1$, $\lambda_2$, and pion isospin indices $a$, $b$. For on-shell particles, the Mandelstam variables defined as
\beq
\label{Mandelstam}
s=(p_1+q_1)^2,\quad t=(q_1-q_2)^2,\quad u=(q_1-p_2)^2,
\eeq
are subject to the constraint
\beq
s+t+u=2\mpi^2.
\eeq
In the center-of-mass frame (CMS), we have
\beq
 t=-2\qq^2(1-z_s),\quad z_s=\cos\theta_s, \quad \qq^2=\frac{(s-\mpi^2)^2}{4s}.
\eeq
The $S$-matrix for the charged process can be written as
\begin{align}
&{}_{\rm out}\langle\gamma(q_2,\lambda_2)\pi^\pm(p_2)|\gamma(q_1,\lambda_1)\pi^\pm(p_1)\rangle_{\rm in}\notag\\
&=(2\pi)^4\delta^4(q_2+p_2-q_1-p_1)\notag\\
&\times\Big\{\delta_{\lambda_1\lambda_2}+ie^2F_{\lambda_1\lambda_2}^{\rm c}(s,t)
e^{i(\lambda_1-\lambda_2)\varphi}\Big\},
\end{align}
with $e^2=4\pi\alpha$ and azimuthal angle $\varphi$, and similarly for the neutral amplitude $F_{\lambda_1\lambda_2}^{\rm n}$. Separating the photon polarization vectors\footnote{Here and below, we suppress isospin indices whenever possible.},
\beq
\label{gauge_inv1}
F_{\lambda_1\lambda_2}(s,t)=\eps_\mu(q_1,\lambda_1)\eps_\nu^*(q_2,\lambda_2)W^{\mu\nu}(s,t),
\eeq
we can use gauge and Lorentz invariance to decompose the amplitude as
\begin{align}
\label{gauge_inv2}
W_{\mu\nu}(s,t)&=A(s,t)\Big(\frac{t}{2}g_{\mu\nu}+q_{2\mu}q_{1\nu}\Big)+B(s,t)\Big(2t\Delta_\mu\Delta_\nu\notag\\
&-(s-u)^2g_{\mu\nu}+2(s-u)(\Delta_\mu q_{1\nu}+\Delta_\nu q_{2\mu})\Big),
\end{align}
where $\Delta_\mu=p_{1\mu}+p_{2\mu}$ and we have dropped terms that vanish in $F_{\lambda_1\lambda_2}$ due to $\eps(q_i,\lambda_i)\cdot q_i=0$. In the conventions of \cite{Edmonds} for the polarization vectors, one obtains
\begin{align}
\label{invar_ampl}
F_{++}(s,t)&=F_{--}(s,t)=4(\mpi^4-s u)B(s,t),\\
F_{+-}(s,t)&=F_{-+}(s,t)=-\frac{t}{2}A(s,t)+t(t-4\mpi^2)B(s,t).\notag
\end{align}
The Born-term contributions are
\begin{align}
\label{born}
A^{\rm Born}(s,t)&\equiv A^{\rm c,\, Born}(s,t)=\frac{1}{\mpi^2-s}+\frac{1}{\mpi^2-u}\notag\\
&=2t B^{\rm c,\, Born}(s,t)\equiv 2t B^{\rm Born}(s,t).
\end{align}
In these conventions, the differential cross section is given by
\beq
\frac{\diff \sigma}{\diff \Omega}=\frac{\alpha^2}{4s}\big(|F_{++}(s,t)|^2+|F_{+-}(s,t)|^2\big).
\eeq

The analytic continuation of $F_{+\pm}(s,t)$ to the kinematical region where $t\geq 4\mpi^2$ describes the crossed-channel process $\gamma\gamma\to\pi\pi$, such that the cross section reads
\begin{align}
\frac{\diff \sigma}{\diff \Omega}\bigg|_{\gamma\gamma\to\pi^+\pi^-}&\hspace{-8pt}=\frac{\alpha^2}{8t}\sigma(t)\big(|F_{++}^{\rm c}(s,t)|^2+|F_{+-}^{\rm c}(s,t)|^2\big),
\end{align}
where $\sigma(t)=\sqrt{1-4 M_\pi^2/t}$. The formula is the same, albeit with
 an additional factor of $1/2$ on the right-hand side, for the case of neutral pions.
It should be noted, though, that the kinematics for the crossed reaction
\beq
\gamma(q_1,\lambda_1)\gamma(-q_2,\lambda_2)\rightarrow \pi^a(-p_1)\pi^b(p_2)
\eeq
in terms of the Mandelstam variables \eqref{Mandelstam} lead to different center-of-mass momenta for initial and final states
\beq
\qt^2=\frac{t}{4},\quad \pt^2=\frac{t}{4}-\mpi^2,
\eeq
and to the CMS scattering angle
\beq
\label{angle_tchannel}
z_t=\cos\theta_t=\frac{\nu}{4p_tq_t},\quad \nu=s-u,\quad p_t=|\pt|, \quad q_t=|\qt|.
\eeq

\subsection{Partial-wave expansion and pion polarizabilities}

The partial-wave expansion of the amplitudes for pion Compton scattering reads \cite{JW}
\beq
\label{PWE_s}
F_{+\pm}(s,t)=\sum\limits_{J=1}^\infty (2J+1)f_{J,\pm}(s)d_{1,\pm1}^J(z_s),
\eeq
with the Wigner $d$-functions\footnote{For convenience, we write $d^J_{m,m'}(\cos \theta)$ instead of $d^J_{m,m'}(\theta)$.}
\beq
d^J_{1,\pm 1}(z)=\frac{1\mp z}{J(J+1)}P_J'(z)\pm P_J(z)
\eeq
 and the inversion
\beq
f_{J,\pm}(s)=\frac{1}{2}\int\limits_{-1}^1\diff z_s d^J_{1,\pm 1}(z_s)F_{+\pm}(s,t)\Big|_{t=-2\qq^2(1-z_s)}.
\eeq
The expansion \eqref{PWE_s} can be mapped onto the multipole expansion \cite{Guiasu78} via\footnote{The covariant amplitudes in~\cite{Guiasu78} are related to ours by $A_{\rm GR}=-e^2 A$, $B_{\rm GR}=16e^2 B$.}
\beq
 f_{J,\pm}(s)=\pm \frac{2\sqrt{s}}{\alpha(2J+1)}(E_J(\omega)\pm M_J(\omega)),
\eeq
where $\omega$ denotes the energy of the photon.
Defining the pion polarizabilities as the leading terms of the Born-term-subtracted multipoles $\hat E_J(\omega)$, $\hat M_J(\omega)$ in an expansion in $\omega$ \cite{Guiasu78},
\begin{align}
\alpha_J&=\frac{2J[(2J-1)!!]^2}{J+1}\frac{\hat E_J(\omega)}{\omega^{2J}}\bigg|_{\omega=0},\notag\\
\beta_J&=\frac{2J[(2J-1)!!]^2}{J+1}\frac{\hat M_J(\omega)}{\omega^{2J}}\bigg|_{\omega=0},
\end{align}
we can read off $\alpha_J$ and $\beta_J$ from an expansion of the Born-term-subtracted amplitudes $\hat F_{+\pm}(s,t)$ in $t$ at fixed $s=\mpi^2$
\begin{align}
\label{pol_exp}
 \frac{2\alpha}{\mpi t}\hat F_{++}(s=\mpi^2,t)&=\alpha_1+\beta_1+\frac{t}{12}(\alpha_2+\beta_2)+\Order(t^2),\notag\\
\frac{-2\alpha}{\mpi t}\hat F_{+-}(s=\mpi^2,t)&=\alpha_1-\beta_1+\frac{t}{12}(\alpha_2-\beta_2)+\Order(t^2).
\end{align}

\subsection{Relation to $\boldsymbol{\gamma\gamma\to\pi\pi}$} 

To establish connection to the notation of the crossed process \cite{GM_10,GIS05}, which we will refer to as the $t$-channel reaction, we briefly discuss
\beq
\gamma(q_1,\lambda_1)\gamma(q_2,\lambda_2)\rightarrow \pi^a(p_1)\pi^b(p_2)
\eeq
in terms of the Mandelstam variables
\beq
\tils=(q_1+q_2)^2,\quad \tilt=(q_1-p_1)^2, \quad \tilu=(q_1-p_2)^2,
\eeq
and the amplitudes
\begin{align}
&{}_{\rm out}\langle \pi(p_1)\pi(p_2)|\gamma(q_1,\lambda_1)\gamma(q_2,\lambda_2)\rangle_{\rm in}\\
&=ie^2(2\pi)^4\delta^4(q_2+p_2-q_1-p_1)H_{\lambda_1\lambda_2}(\tils,\tilt)\notag
e^{i(\lambda_1-\lambda_2)\varphi}.
\end{align}
They are related to the $s$-channel amplitudes by
\beq
\label{schannel_tchannel_rel}
H_{++}(s,t)=- F_{+-}(t,s),\quad H_{+-}(s,t)=- F_{++}(t,s),
\eeq
and their polarizability expansion therefore reads 
\begin{align}
 \frac{2\alpha}{\mpi \tils}\hat H_{++}(\tils,\tilt=\mpi^2)&=\alpha_1-\beta_1+\frac{\tils}{12}(\alpha_2-\beta_2)+\Order(\tils^2),\notag\\
\frac{-2\alpha}{\mpi \tils}\hat H_{+-}(\tils,\tilt=\mpi^2)&=\alpha_1+\beta_1+\frac{\tils}{12}(\alpha_2+\beta_2)+\Order(\tils^2).
\end{align}
Furthermore, the partial-wave amplitudes $h_{J \pm}$ follow from $F_{+ \pm}$ via
\begin{align}
\label{PWE_t}
F_{++}(s,t)&=-\sum\limits_{J}(2J+1)h_{J,-}(t)d^J_{20}(z_t),\notag\\
F_{+-}(s,t)&=-\sum\limits_{J}(2J+1)h_{J,+}(t)d^J_{00}(z_t),
\end{align}
where due to Bose symmetry the sum extends over even values of $J$ only, and 
\beq
d^J_{00}(z)=P_J(z),\quad d^J_{20}(z)=\frac{2P'_{J-1}(z)-J(J-1)P_J(z)}{\sqrt{(J-1)J(J+1)(J+2)}}.
\eeq
In our conventions, the transition between isospin and particle basis is achieved by
\beq
\label{isospin}
\begin{pmatrix}
 H^{\rm c}\\ H^{\rm n}
\end{pmatrix}
=\begin{pmatrix}
  \frac{1}{\sqrt{3}} & \frac{1}{\sqrt{6}}\\
\frac{1}{\sqrt{3}} & -\sqrt{\frac{2}{3}}
 \end{pmatrix}
\begin{pmatrix}
 H^0\\ H^2
\end{pmatrix}.
\eeq

\section{Roy--Steiner equations}
\label{sec:roy_steiner}

First we review the basic steps in the construction of Roy equations in $\pi\pi$ scattering. They are \cite{Roy}:
\begin{enumerate}
 \item Write down a twice-subtracted dispersion relation at fixed Mandelstam $t$, whose subtraction ``constants'' are actually functions which depend on the value of $t$ chosen.
\item Use crossing symmetry to determine the subtraction functions, such that the remaining free parameters are the $\pi\pi$ scattering lengths.
\item Expand the imaginary part of the amplitude that appears under the dispersion integrals in partial waves, and perform a partial-wave projection of the resulting equation.
\end{enumerate}
In this way, one arrives at a system of integral equations for the $\pi\pi$ amplitudes $t_l^I(s)$
\beq
t_l^I(s)=k_l^I(s)+\sum\limits_{I'=0}^{2}\sum\limits_{l'=0}^\infty\int\limits_{4M_\pi^2}^\infty \text{d}s'K_{ll'}^{II'}(s,s')\text{Im}\,t_{l'}^{I'}(s')\label{Royeq},
\eeq
that relates a partial wave of a given angular momentum $l$ and isospin $I$ to all other partial waves via analytically calculable kinematic kernel functions $K_{ll'}^{II'}(s,s')$. A detailed discussion of how to numerically solve this system can be found in \cite{ACGL}.

\begin{sloppypar}
The construction of Roy equations for reactions with non-identical particles is hampered by the fact that crossing symmetry intertwines different physical processes, such that the second step based solely on fixed-$t$ dispersion relations fails. For this reason, hyperbolic dispersion relations \cite{HS_73} were used in \cite{AB_00,piK} to determine the subtraction functions and derive Roy equations for $\pi K$ scattering, as within the hyperbolic approach the dispersion relations automatically involve  both $\pi K\to \pi K$ and $\pi\pi\to K \bar K$ physical regions. The resulting system in such a case is therefore referred to as Roy--Steiner equations.
\end{sloppypar}

In this work, we will go a step further and solely consider hyperbolic dispersion relations. This is particularly convenient because $s\leftrightarrow u$ crossing symmetry is manifest, and both $\gamma\pi\to\gamma\pi$ and $\gamma\gamma\to\pi\pi$ amplitudes contribute, such that all constraints by crossing symmetry are automatically fulfilled. We now turn to the setup of this system of Roy--Steiner equations.

\subsection{Hyperbolic dispersion relations}

We start by writing down unsubtracted hyperbolic dispersion relation for the amplitudes $A$ and $B$, which can be constructed following \cite{HS_73}. The advantage of using dispersion relations for $A$ and $B$ instead of $F_{+\pm}$ is that all constraints by gauge invariance that lead to the decomposition \eqref{invar_ampl} are automatically built in. In particular, the equations for $F_{++}$ and $F_{+-}$ do not decouple, as gauge invariance dictates that the same invariant function $B$ contributes to both amplitudes. The dispersion relations for $A$ and $B$ read
\begin{align}
\label{hyp_disp}
 A(s,t)&=\bar A^{\rm B}(s,t)
+\frac{1}{\pi}\int\limits_{4\mpi^2}^\infty\diff t'\frac{\Im A(t',z_t')}{t'-t}\notag\\
&+\frac{1}{\pi}\int\limits_{\mpi^2}^\infty\diff s'\Im A(s',t')\bigg(\frac{1}{s'-s}+\frac{1}{s'-u}-\frac{1}{s'-a}\bigg),\notag\\
 B(s,t)&=\bar B^{\rm B}(s,t)
+\frac{1}{\pi}\int\limits_{4\mpi^2}^\infty\diff t'\frac{\Im B(t',z_t')}{t'-t}\\
&+\frac{1}{\pi}\int\limits_{\mpi^2}^\infty\diff s'\Im B(s',t')\bigg(\frac{1}{s'-s}+\frac{1}{s'-u}-\frac{1}{s'-a}\bigg),\notag
\end{align}
with the Born terms 
\beq
 \bar A^{\rm B}(s,t)=A^{\rm Born}(s,t)-\frac{1}{\mpi^2-a},\quad
\bar B^{\rm B}(s,t)=B^{\rm Born}(s,t)
\eeq
only contributing to the charged-pion process ($A^{\rm Born}$ and $B^{\rm Born}$ were already defined in~\eqref{born}). The primed set of Mandelstam variables is constrained to lie on the hyperbola
\beq
\label{hyp}
(s'-a)(u'-a)=(s-a)(u-a)\equiv b,
\eeq
where the hyperbola parameter $a$ can be freely chosen. In particular, it can be used to optimize the range of validity of the resulting system of Roy--Steiner equations (cf.~Sect.~\ref{sec:dom_val}). The above integrals are understood such that the integrands shall be expressed in terms of the integration variable and the external kinematics by virtue of \eqref{hyp} and
\beq
s'+t'+u'=2\mpi^2.
\eeq
The second integral in \eqref{hyp_disp} is reminiscent of fixed-$t$ dispersion relations, but in that case $\Im A(s',t') \to  \Im A(s',t)$ and the last term is removed. Thus, the key difference here is that $t'$ depends not only on $t$, but on $s$ and $s'$ as well. However, it is possible to recover the limit of fixed-$t$ by sending $a$ to infinity. This can be shown explicitly based on the relation between $z_s$ and $z_s'$ given in \eqref{zs_zsprime}.

\subsection{Sum rules and subtracted dispersion relations}

The most economical way to obtain a subtracted version of hyperbolic dispersion relations is to derive sum rules from the original dispersion relation and then subtract them from it. We will choose the subtraction points such that the subtraction constants coincide with the pion polarizabilities. 

With knowledge of the dipole pion polarizabilities $\alpha_1\pm \beta_1$ we can implement one subtraction, while for a second subtraction the quadrupole polarizabilities $\alpha_2\pm\beta_2$ are also needed.
For example, choosing $s=M_\pi^2$ and taking the limit $t \rightarrow 0$, we can compare \eqref{hyp_disp} with \eqref{invar_ampl} and \eqref{pol_exp} in order to obtain\footnote{Note that $z_t'$ and $t'$ depend on the integration variable as well as on $s$ and $t$. The subscript $0$ indicates evaluation at $s=\mpi^2$ and $t=0$.}
\begin{align}
\frac{\mpi}{2\alpha}(\alpha_1+\beta_1)&=\frac{4\mpi^2}{\pi}\int\limits_{4\mpi^2}^\infty\diff t'\frac{\Im B(t',z_t')\big|_0}{t'}\\
&\hspace{-25pt}+\frac{4\mpi^2}{\pi}\int\limits_{\mpi^2}^\infty\diff s'\Im B(s',t')\big|_0\bigg(\frac{2}{s'-\mpi^2}-\frac{1}{s'-a}\bigg),\notag
\end{align}
which, together with similar sum rules, may be used to write down subtracted versions of \eqref{hyp_disp}. 

\subsection{$\boldsymbol{s}$-channel projection}

The Roy--Steiner system is obtained by expanding the integrands into partial waves and subsequently projecting  each equation onto $s$- and $t$-channel partial waves. To achieve this we use \eqref{PWE_s} and \eqref{PWE_t} as well as their inversions. It is useful to note that
\beq
4(\mpi^4-s u)=8s \qq^2(1+z_s)=-t(t-4\mpi^2)(1-z_t^2)
\eeq
to identify the relevant kinematic dependencies.

We start with the projection onto $\gamma\pi\to\gamma\pi$ partial waves, which can be written as
\begin{align}
\label{schanneleq}
f_{J,+}(s)&=N_J^+(s)+\frac{1}{\pi}\int\limits_{\mpi^2}^\infty\diff s'\sum\limits_{J'=1}^\infty K_{JJ'}^{++}(s,s')\Im f_{J',+}(s')\notag\\
&+\frac{1}{\pi}\int\limits_{4\mpi^2}^\infty\diff t'\sum_{J'\,{\rm even}} G_{JJ'}^{+-}(s,t')\Im h_{J',-}(t'),\notag\\
f_{J,-}(s)&=N_J^-(s)+\frac{1}{\pi}\int\limits_{\mpi^2}^\infty\diff s'\sum\limits_{J'=1}^\infty \Big(K_{JJ'}^{-+}(s,s')\Im f_{J',+}(s')\notag\\
&\qquad+K_{JJ'}^{--}(s,s')\Im f_{J',-}(s')\Big)\notag\\
&+\frac{1}{\pi}\int\limits_{4\mpi^2}^\infty\diff t'\sum_{J'\,{\rm even}} \Big(G_{JJ'}^{-+}(s,t')\Im h_{J',+}(t')\notag\\
&\qquad+G_{JJ'}^{--}(s,t')\Im h_{J',-}(t')\Big),
\end{align}
where $N_J^\pm(s)$ includes Born terms and---in case subtractions were performed---pion polarizabilities. The kernel functions for the unsubtracted case read
\begin{align}
K_{JJ'}^{++}(s,s')&=\frac{s\, \qq^2}{s'\qq'^2}\frac{2J'+1}{2}\int\limits_{-1}^1\diff z_s(1+z_s)d^J_{11}(z_s)\notag\\
&\times\frac{d^{J'}_{11}(z_s')}{1+z_s'}
\bigg\{\frac{1}{s'-s}+\frac{1}{s'-u}-\frac{1}{s'-a}\bigg\},\notag\\
G_{JJ'}^{+-}(s,t')&=\frac{8s\qq^2}{t'(t'-4\mpi^2)}\frac{2J'+1}{2}\int\limits_{-1}^1\diff z_s(1+z_s)d^J_{11}(z_s)\notag\\
&\times\frac{d^{J'}_{20}(z_t')}{1-z_t'^2}\frac{1}{t'-t},\notag\\
K_{JJ'}^{-+}(s,s')&=\frac{\qq^2}{4s'\qq'^2}\frac{2J'+1}{2}\int\limits_{-1}^1\diff z_s(1-z_s)d^J_{1,-1}(z_s)\notag\\
&\hspace{-20pt}\times\frac{d^{J'}_{11}(z_s')}{1+z_s'}(t'-t)
\bigg\{\frac{1}{s'-s}+\frac{1}{s'-u}-\frac{1}{s'-a}\bigg\},\notag\\
K_{JJ'}^{--}(s,s')&=\frac{\qq^2}{\qq'^2}\frac{2J'+1}{2}\int\limits_{-1}^1\diff z_s(1-z_s)d^J_{1,-1}(z_s)\notag\\
&\times\frac{d^{J'}_{1,-1}(z_s')}{1-z_s'}
\bigg\{\frac{1}{s'-s}+\frac{1}{s'-u}-\frac{1}{s'-a}\bigg\},\notag\\
G_{JJ'}^{-+}(s,t')&=2\qq^2\frac{2J'+1}{2}\int\limits_{-1}^1\diff z_s\frac{1-z_s}{t'(t'-t)}d^J_{1,-1}(z_s)P_{J'}(z_t'),\notag\\
G_{JJ'}^{--}(s,t')&=2\qq^2\frac{2J'+1}{2}\int\limits_{-1}^1\diff z_s\frac{1-z_s}{t'(t'-4\mpi^2)}d^J_{1,-1}(z_s)\notag\\
&\times\frac{d^{J'}_{20}(z_t')}{1-z_t'^2}.
\end{align}
Explicit expressions for $J,J'\leq 2$ as well as the modifications for the subtracted case are given in App.~\ref{app:kernel_schannel}.
It is important to note that while we have consistently suppressed isospin indices for the partial-wave amplitudes, all kernel functions are independent of isospin.

\subsection{$\boldsymbol{t}$-channel projection}

Similarly, the projection onto $\gamma\gamma\to\pi\pi$ amplitudes has the form
\begin{align}
\label{tchanneleq}
h_{J,+}(t)&=\tilde N_J^+(t)+\frac{1}{\pi}\int\limits_{\mpi^2}^\infty\diff s'\sum\limits_{J'=1}^\infty \Big(\tilde G_{JJ'}^{++}(t,s')\Im f_{J',+}(s')\notag\\
&\qquad+\tilde G_{JJ'}^{+-}(t,s')\Im f_{J',-}(s')\Big)\notag\\
&+\frac{1}{\pi}\int\limits_{4\mpi^2}^\infty\diff t'\sum_{J'\,{\rm even}} \Big(\tilde K_{JJ'}^{++}(t,t')\Im h_{J',+}(t')\notag\\
&\qquad+\tilde K_{JJ'}^{+-}(t,t')\Im h_{J',-}(t')\Big),\notag\\
h_{J,-}(t)&=\tilde N_J^-(t)+\frac{1}{\pi}\int\limits_{\mpi^2}^\infty\diff s'\sum\limits_{J'=1}^\infty \tilde G_{JJ'}^{-+}(t,s')\Im f_{J',+}(s')\notag\\
&+\frac{1}{\pi}\int\limits_{4\mpi^2}^\infty\diff t'\sum_{J'\,{\rm even}} \tilde K_{JJ'}^{--}(t,t')\Im h_{J',-}(t'),
\end{align}
where, in the unsubtracted case,
\begin{align}
\tilde G_{JJ'}^{++}(t,s')&=\frac{t}{8s'\qq'^2}\frac{2J'+1}{2}\int\limits_{-1}^1\diff z_t(t'-t)P_J(z_t)\notag\\
&\times\frac{d_{11}^{J'}(z_s')}{1+z_s'}\bigg\{\frac{1}{s'-s}+\frac{1}{s'-u}-\frac{1}{s'-a}\bigg\},\notag\\
\tilde G_{JJ'}^{+-}(t,s')&=\frac{t}{2\qq'^2}\frac{2J'+1}{2}\int\limits_{-1}^1\diff z_tP_J(z_t)\notag\\
&\times\frac{d_{1,-1}^{J'}(z_s')}{1-z_s'}\bigg\{\frac{1}{s'-s}+\frac{1}{s'-u}-\frac{1}{s'-a}\bigg\},\notag\\
\tilde K_{JJ'}^{++}(t,t')&=\frac{t}{t'(t'-t)}\frac{2J'+1}{2}\int\limits_{-1}^1\diff z_t P_J(z_t)P_{J'}(z_t'),\notag\\
\tilde K_{JJ'}^{+-}(t,t')&=\frac{t}{t'(t'-4\mpi^2)}\frac{2J'+1}{2}\int\limits_{-1}^1\diff z_t P_J(z_t)\frac{d^{J'}_{20}(z_t')}{1-z_t'^2},\notag\\
\tilde G_{JJ'}^{-+}(t,s')&=\frac{t(t-4\mpi^2)}{8s'\qq'^2}\frac{2J'+1}{2}\int\limits_{-1}^1\diff z_t(1-z_t^2)d_{20}^J(z_t)\notag\\
&\times\frac{d_{11}^{J'}(z_s')}{1+z_s'}\bigg\{\frac{1}{s'-s}+\frac{1}{s'-u}-\frac{1}{s'-a}\bigg\},\notag\\
\tilde K_{JJ'}^{--}(t,t')&=\frac{t(t-4\mpi^2)}{t'(t'-4\mpi^2)(t'-t)}\\
&\times\frac{2J'+1}{2}\int\limits_{-1}^1\diff z_t (1-z_t^2) d_{20}^J(z_t)\frac{d_{20}^{J'}(z_t')}{1-z_t'^2}.\notag
\end{align}
Explicit expressions for $J,J'\leq 2$ are provided in App.~\ref{app:kernel_tchannel}.

\subsection{Threshold and asymptotic behavior of the kernel functions}

\begin{table}
\centering
\begin{tabular}{cccc}
\hline\hline
 \# subtractions& $0$ & $1$ & $2$ \\\hline
$K_{11}^{++}(s,s'), K_{12}^{++}(s,s')$ & $\Order(s'^{-3})$ & $\Order(s'^{-4})$ & $\Order(s'^{-5})$ \\
$K_{21}^{++}(s,s'), K_{22}^{++}(s,s')$ & $\Order(s'^{-4})$ & $\Order(s'^{-4})$ & $\Order(s'^{-5})$ \\
$G_{12}^{+-}(s,t')$ & $\Order(t'^{-3})$ & $\Order(t'^{-4})$ & $\Order(t'^{-5})$ \\
$G_{22}^{+-}(s,t')$ & $\Order(t'^{-4})$ & $\Order(t'^{-4})$ & $\Order(t'^{-5})$ \\
$\tilde G_{21}^{-+}(t,s'), \tilde G_{22}^{-+}(t,s')$ & $\Order(s'^{-3})$ & $\Order(s'^{-4})$ & $\Order(s'^{-5})$ \\
$\tilde K^{--}_{22}(t,t')$ & $\Order(t'^{-3})$ & $\Order(t'^{-4})$ & $\Order(t'^{-5})$ \\
\hline\hline
\end{tabular}
\caption{Asymptotics of the kernel functions in the equations for $f_{J,+}(s)$ and $h_{J,-}(t)$.}
\label{table:asym_fpp}
\end{table}

\begin{table}
\centering
\begin{tabular}{ccccc}
\hline\hline
\# subtractions& $0$ & $1$ & $2$  \\\hline
$K_{11}^{-+}(s,s'), K_{12}^{-+}(s,s')$ & $\Order(s'^{-2})$ & $\Order(s'^{-3})$ & $\Order(s'^{-4})$  \\
$K_{21}^{-+}(s,s'), K_{22}^{-+}(s,s')$ & $\Order(s'^{-4})$ & $\Order(s'^{-3})$ & $\Order(s'^{-4})$  \\
$K_{11}^{--}(s,s'), K_{12}^{--}(s,s')$ & $\Order(s'^{-2})$ & $\Order(s'^{-3})$ & $\Order(s'^{-4})$  \\
$K_{21}^{--}(s,s'), K_{22}^{--}(s,s')$ & $\Order(s'^{-3})$ & $\Order(s'^{-3})$ & $\Order(s'^{-4})$  \\
$G_{10}^{-+}(s,t'), G_{12}^{-+}(s,t')$ & $\Order(t'^{-2})$ & $\Order(t'^{-3})$ & $\Order(t'^{-4})$ \\
$G_{20}^{-+}(s,t'), G_{22}^{-+}(s,t')$ & $\Order(t'^{-3})$ & $\Order(t'^{-3})$ & $\Order(t'^{-4})$ \\
$G_{12}^{--}(s,t')$ & $\Order(t'^{-2})$ & $\Order(t'^{-3})$ & $\Order(t'^{-4})$ \\
$G_{22}^{--}(s,t')$ & --- & $\Order(t'^{-3})$ & $\Order(t'^{-4})$ \\
$\tilde G_{01}^{+\pm}(t,s'), \tilde G_{02}^{+\pm}(t,s')$ & $\Order(s'^{-2})$ & $\Order(s'^{-3})$ & $\Order(s'^{-4})$  \\
$\tilde G_{21}^{+\pm}(t,s'), \tilde G_{22}^{+\pm}(t,s')$ & $\Order(s'^{-4})$ & $\Order(s'^{-4})$ & $\Order(s'^{-4})$  \\
$\tilde K^{++}_{00}(t,t'), \tilde K^{++}_{02}(t,t')$ & $\Order(t'^{-2})$ & $\Order(t'^{-3})$ & $\Order(t'^{-4})$ \\
$\tilde K^{++}_{22}(t,t')$ & $\Order(t'^{-4})$ & $\Order(t'^{-4})$ & $\Order(t'^{-4})$\\
$\tilde K^{+-}_{02}(t,t')$ & $\Order(t'^{-2})$ & $\Order(t'^{-3})$ & $\Order(t'^{-4})$\\
\hline\hline
\end{tabular}
\caption{Asymptotics of the kernel functions in the equations for $f_{J,-}(s)$ and $h_{J,+}(t)$.}
\label{table:asym_fpm}
\end{table}

\begin{sloppypar}
In order to check our kernel functions and determine the convergence properties of the dispersive integrals, we study the behavior of the kernels at threshold and for $s', t'\to\infty$.
\end{sloppypar}

Based on \eqref{PWE_t}, one can show that for $t\to 0$
\begin{align}
\label{threshold_h}
\hat H_{++}(t,\mpi^2)=\hat h_{0,+}(t)-\frac{5}{2}\hat h_{2,+}(t)+\Order(t^3),\notag\\
\hat H_{+-}(t,\mpi^2)=-\frac{5\sqrt{6}\,\mpi^2}{t-4\mpi^2}\hat h_{2,-}(t)+\Order(t^2),
\end{align}
where the hat indicates the subtraction of the Born terms.
By comparing to \eqref{pol_exp} and \eqref{tchannel_Born}, \eqref{threshold_h} implies that
\begin{align}
 \tilde G_{JJ'}^{+\pm,\,(1)}(t,s')&=\Order(t^2),\quad \tilde K_{JJ'}^{+\pm,\,(1)}(t,t')=\Order(t^2),\notag\\
\tilde G_{0J'}^{+\pm,\,(2)}(t,s')&-\frac{5}{2}\tilde G_{2J'}^{+\pm,\,(2)}(t,s')=\Order(t^3),\notag\\
\tilde K_{0J'}^{+\pm,\,(2)}(t,t')&-\frac{5}{2}\tilde K_{2J'}^{+\pm,\,(2)}(t,t')=\Order(t^3),\notag\\
\tilde G_{2J'}^{-+,\,(1)}(t,s')&=\Order(t^2),\quad \tilde G_{2J'}^{-+,\,(2)}(t,s')=\Order(t^2),\notag\\
\tilde K_{2J'}^{--,\,(1)}(t,t')&=\Order(t^2),\quad \tilde K_{2J'}^{--,\,(2)}(t,t')=\Order(t^2),
\end{align}
where the superscript refers to the number of subtractions.
We have checked that the explicit expressions in App.~\ref{app:kernel_tchannel} fulfill these relations.

Furthermore, is clear from \eqref{schanneleq} and \eqref{tchanneleq} that the dependence of the kernel functions on $s$ and $t$ must reproduce the correct threshold behavior of the partial-wave amplitudes
\begin{align}
 f_{J,+}(s)&=\Order\big(\qq^{2J}\big),\quad f_{J,-}(s)=\Order\big(\qq^{2J}\big),\notag\\
\hat h_{J,+}(t)&=\Order\big(q_t^2(q_tp_t)^J\big),\quad \hat h_{J,-}(t)=\Order\big((q_tp_t)^J\big).
\end{align}
 The additional factor of $q_t^2$ in $\hat h_{J,+}(t)$ is a manifestation of Low's theorem for low-energy QED \cite{Low}, which requires the full scattering amplitude to be equal to the Born terms at the threshold for Compton scattering. We have checked explicitly that the expressions provided in Apps.~\ref{app:kernel_schannel} and \ref{app:kernel_tchannel} indeed fulfill
\begin{align}
 K_{JJ'}^{++}(s,s')&=\Order\big(\qq^{2J}\big),\quad G^{+-}_{JJ'}(s,t')=\Order\big(\qq^{2J}\big),\notag\\
K_{JJ'}^{-\pm}(s,s')&=\Order\big(\qq^{2J}\big),\quad G^{-\pm}_{JJ'}(s,t')=\Order\big(\qq^{2J}\big),\\
\tilde G_{JJ'}^{+\pm}(t,s')&=\Order\big(q_t^2(q_tp_t)^J\big),\quad \tilde K_{JJ'}^{+\pm}(t,t')=\Order\big(q_t^2(q_tp_t)^J\big),\notag\\
\tilde G_{JJ'}^{-+}(t,s')&=\Order\big((q_tp_t)^J\big),\quad \tilde K_{JJ'}^{--}(t,t')=\Order\big((q_tp_t)^J\big).\notag
\end{align}

Similarly, the asymptotic behavior of the kernel functions for $s'\to\infty$ and $t'\to\infty$, respectively, determines the convergence properties of the dispersion integrals. In particular, one can directly read off which rate of convergence can be achieved when working with a certain number of subtractions.
The corresponding behavior of the kernels for large values of the respective integration variable is summarized in Tables \ref{table:asym_fpp} and \ref{table:asym_fpm}. Although in some cases the leading power vanishes, one can see that in general the kernels for $f_{J,-}(s)$ and $h_{J,+}(t)$ drop as the second, third, and fourth power in the integration variable, while the integrals related to $f_{J,+}(s)$ and $h_{J,-}(t)$ will converge one order faster.

\section{Domain of validity}
\label{sec:dom_val}

The original Roy equations for $\pi\pi$ scattering have been shown to be rigorously valid up to $s_{\rm max}=60\mpi^2$ based on axiomatic field theory \cite{Roy}. This range can be extended further to $s_{\rm max}=68\mpi^2$ if the $\pi \pi$ amplitude ${\cal T}$ fulfills Mandelstam analyticity \cite{Mandelstam58}, i.e.~it can be written in terms of double spectral functions $\rho_{su}$, $\rho_{tu}$, and $\rho_{st}$ as
\begin{align}
\label{mandelstam}
{\cal T}(s,t)&=\frac{1}{\pi^2}\iint\diff s'\diff u'\frac{\rho_{su}(s',u')}{(s'-s)(u'-u)}\notag\\
&+ \frac{1}{\pi^2}\iint\diff t'\diff u'\frac{\rho_{tu}(t',u')}{(t'-t)(u'-u)}\notag\\
&+\frac{1}{\pi^2}\iint\diff s'\diff t'\frac{\rho_{st}(s',t')}{(s'-s)(t'-t)}.
\end{align}
In this section, we derive the corresponding limits on our system of $\gamma\pi\to\gamma\pi$ Roy--Steiner equations. For more details of the derivation we refer to similar work on $\pi K$ and $\pi N$ scattering \cite{HS_73,piK,Hoehler}.

\begin{figure}
\centering
\includegraphics[height=\linewidth,angle=-90]{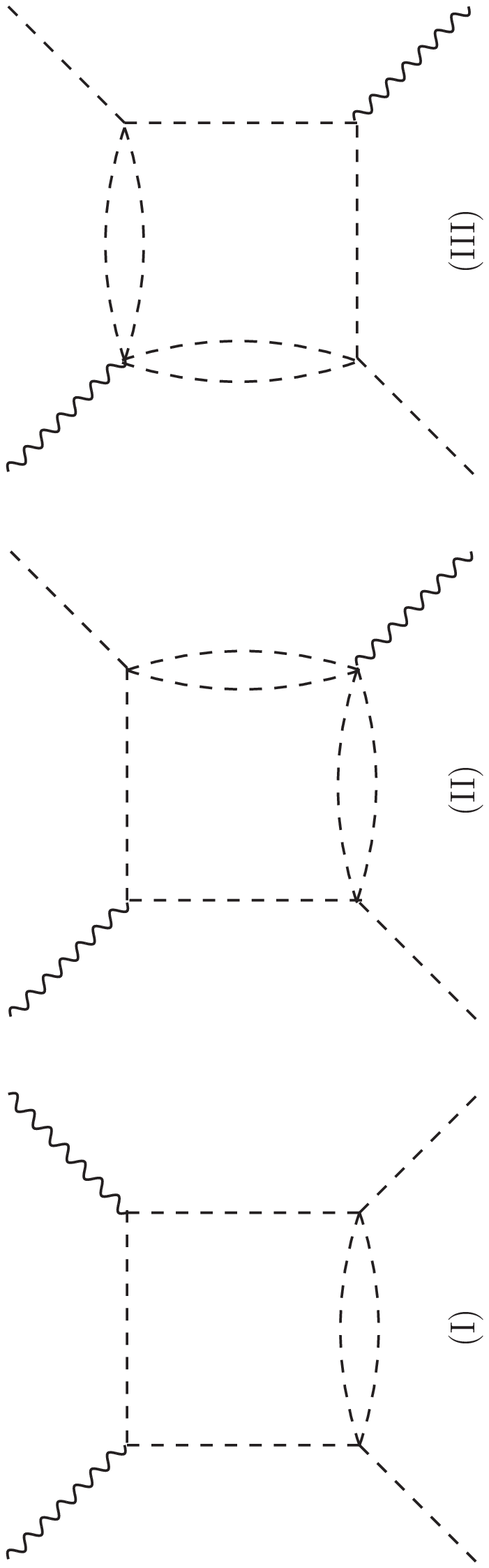}
\caption{Box graphs constraining the boundaries of the double spectral functions.}
\label{fig:boundary}
\end{figure}

The central objects of the discussion will be the boundaries of the support of the double spectral functions, which determine the integration range in \eqref{mandelstam}. These boundaries can be inferred from the box diagrams depicted in Fig.\  \ref{fig:boundary}. These diagrams are to be understood as generalizations of four-propagator box diagrams~\cite{IZ}, with one or more lines replaced by a particle whose mass is equal to the input mass of the lowest-lying intermediate state accessible to the interacting particles. If we neglect the possibility of photons in intermediate states, the pertinent states involve one and two pions. 
The diagram $({\rm I})$ represents the mechanism that produces the boundary of the support of $\rho_{st}$, while $({\rm II})$ and $({\rm III})$  are relevant for $\rho_{su}$. ``Double-spectral regions" are then defined as the portions of the Mandelstam plane that obey $s + t + u=2 M_\pi^2$ and where {\it any} of the functions $\rho_{st}$, $\rho_{su}$, $\rho_{tu}$ has support. 

From (I) we find that
\beq
 b_{\rm I}(s,t)=t(s-9\mpi^2)-4\mpi^2(s-\mpi^2)=0\label{st}
\eeq
defines the $st$ boundary of support, and the same with $s\leftrightarrow u$ for the $tu$ boundary, while
\beq
 b_{\rm II}(s,t)=b_{\rm III}(s,t)=s u+\mpi^2(9t-\mpi^2)=0
\eeq
is the boundary of $\rho_{su}$. In particular, the boundary of $\rho_{st}$ may be described by
\beq
t=T_{st}(s)=T_{\rm I}(s)\label{Tst},
\eeq
where $t=T_{\rm I}(s)$ follows from solving $b_{\rm I}(s,t)=0$ for $t$, and similarly for $\rho_{tu}$ and $\rho_{su}$.

These domains of support restrict the range of validity of the Roy--Steiner equations  
in two ways:
\begin{enumerate}
\item The partial-wave expansion of the imaginary parts in the dispersion integrals converges only for scattering angles $z$ that lie within the large Lehmann ellipse \cite{Lehmann58}. This ellipse can be constructed as the largest ellipse in the complex $z$-plane which does not reach into the double spectral regions. Given a value of $a$, this constraint can be translated into an allowed range for the parameter $b$, since~\eqref{hyp} relates $b$ to the angle $z$.
\item A specific value of $b$ is only allowed if the hyperbola $(s-a)(u-a)=b$ does not enter the double spectral regions.
\end{enumerate}
The allowed values of $b$ (for a given $a$) must respect both constraints and must do so in the integrals which occur in expressions for both the $\gamma \pi \rightarrow \gamma \pi$ and $\gamma \gamma \rightarrow \pi \pi$ amplitudes. Once we have a constraint on the allowed values of $b$ we can translate it into limits of the domain of validity of the full Roy--Steiner system because of the partial-wave projection of the dispersion relations. For example, in the $s$-channel, we need $-1\leq z_s\leq 1$, such that the maximally allowed $s_{\rm max}$ is the largest value of $s$ for which both
\beq
(s-a)(2\mpi^2-s-a)
\eeq 
and
\beq
(s-a)\bigg(2\mpi^2-s-a+\frac{(s-\mpi^2)^2}{s}\bigg)
\eeq
lie within the allowed range for $b$. In a similar fashion the requirement that $0\leq z_t^2\leq 1$ determines $t_{\rm max}$ in the $t$-channel projection. 

We begin with the Lehmann-ellipse constraint in the $s$-channel. The partial-wave expansion converges in an ellipse with foci at $z_s'=\pm 1$
\beq                                                                                                                   
\frac{(\Re z_s')^2}{A_s^2}+\frac{(\Im z_s')^2}{B_s^2}=1,
\eeq
so that the semimajor and semiminor axis $A_s$ and $B_s$ are related by
\beq
A^2_s-B^2_s=1.
\eeq
The maximal value of $z_{s}'$ that does not enter the region where $\rho_{st} \neq 0$ is given by
\beq
z_{s'}^{\rm max}=1+\frac{2s'T_{st}(s')}{(s'-\mpi^2)^2},
\eeq
such that the $st$ boundary provides the Lehmann-ellipse constraint
\beq
-z_{s'}^{\rm max}\leq z_s'\leq z_{s'}^{\rm max}.
\eeq
Translating this into a restriction on $t'$ we find
\begin{align}
\label{tpconst}
T'_{st}(s')&\leq t' \leq T_{st}(s'), \notag\\
T'_{st}(s')&=-\frac{(s'-\mpi^2)^2}{s'}-T_{st}(s'),\quad \forall\, s'\in\big[\mpi^2,\infty\big),
\end{align}
in the $s$-channel integral of the dispersion relations. As \eqref{hyp} defines a linear relation between $b$ and $t'$, \eqref{tpconst} translates into a condition on $b$
\beq
b^-_s(s',a)\leq b \leq b^+_s(s',a), \label{s_channel_Lehmann}
\eeq
where
\begin{align}
b^+_s(s',a)&=(s'-a)(2\mpi^2-s'-T_{st}'(s')-a),\notag\\
b^-_s(s',a)&=(s'-a)(2\mpi^2-s'-T_{st}(s')-a).
\end{align}
We may then define
\begin{align}
b_s^{+}(a)&=\min_{s' \in[\mpi^2,\infty)} b_s^{+}(s',a),\notag\\
b_s^{-}(a)&=\max_{s' \in[\mpi^2,\infty)} b_s^{-}(s',a)
\end{align}
to be the minimum/maximum of $b_s^{\pm}(s',a)$ within the domain of integration. 
Similar restrictions are provided by $\rho_{tu}$ and $\rho_{su}$, and the intersection of the resulting constraints on $b$ defines the limitations imposed by condition~1 due to the $s$-channel part of the dispersion relation. 

The same condition in the case of the $t$-channel reaction involves a slightly more complicated argument,  because
in this case the relation between the CMS angle $z_t'$ and $b$ is not linear
\beq
z_t'^2=\frac{(t'-2\mpi^2+2a)^2-4b}{16q_t'^2p_t'^2}.
\eeq
Consequently, we need to consider the Lehmann ellipse for $z_t'^2$
\beq
\frac{\big(\Re z_t'^2-\frac{1}{2}\big)^2}{\tilde A_t^2}+\frac{\big(\Im z_t'^2\big)^2}{\tilde B_t^2}=1,
\eeq
where the parameters are related to those of the ellipse for $z_t'$ by
\beq
\tilde A_t=A_t^2-\frac{1}{2},\quad \tilde B_t=A_t B_t.
\eeq
Rewriting \eqref{st} in terms of $\nu$ and $t$ and inserting the result in \eqref{angle_tchannel},  we obtain
the boundary of the double-spectral region in terms of $z_t'$
\beq
z_{t'}^{\rm max}=\frac{N(t')}{4q_t'p_t'},\quad N(t')=\frac{t'(t'+12\mpi^2)}{t'-4\mpi^2},
\eeq
such that $\rho_{st}$ imposes the restriction
\beq
b^-_t(t',a)\leq b \leq b^+_t(t',a), \quad \forall\, t'\in\big[4\mpi^2,\infty\big),\label{t_channel_Lehmann}
\eeq
with
\begin{align}
b^-_t(t',a)&=\frac{1}{4}(t'-2\mpi^2+2a)^2-\frac{1}{4}N(t')^2,\notag\\
b^+_t(t',a)&=\frac{1}{4}(t'-2\mpi^2+2a)^2+\frac{1}{4}N(t')^2-4q_t'^2p_t'^2,
\end{align}
and similarly for $\rho_{tu}$ and $\rho_{su}$. Defining the variables $b_t^{\pm}(a)$
\begin{align}
b_t^{+}(a)&=\min_{t' \in[4\mpi^2,\infty)} b_t^{+}(t',a),\notag\\
b_t^{-}(a)&=\max_{t' \in[4\mpi^2,\infty)} b_t^{-}(t',a),
\end{align}
the constraints
\beq
b_s^-(a) \leq b \leq b_s^+(a),\quad
b_t^-(a) \leq b \leq b_t^+(a)
\eeq
then provide, for a given $a$, the range of allowed values for $b$ that satisfies the Lehmann-ellipse constraint for both the $s$-channel and $t$-channel integrals. 

If the associated hyperbolae in the external variables $s$ and $u$, $(s-a)(u-a)=b$, do not cross the double spectral regions either, then this determines the kinematic regime in which the partial-wave projection is valid. 

\subsection{$\boldsymbol{s}$-channel projection}
\label{sec:sconvergence}

The freedom in the choice of the parameter $a$ in the construction of the hyperbolic dispersion relations can be used in order to maximize the region in $s$ within which these dispersion relations are valid.

As a first step, we consider the restrictions due to $\rho_{st}$ in the $s$-channel part only. The strategy to find the optimal value of $a$ goes as follows: the Lehmann-ellipse constraint imposes the condition that all allowed values of $b$ must fulfill $b^-_s(a)\leq b\leq b^+_s(a)$. As $t$ varies only within
\beq
-\frac{(s-\mpi^2)^2}{s}\leq t\leq 0,
\eeq
this limits the range of values of $b$ that are needed for a given $s$. The value of $s$ which is the maximum one possible, $s_{\rm max}$, will be that which yields a result where the smallest $b$ coincides with $b_s^-(a)$, and the largest $b$ coincides with $b_s^+(a)$, i.e. the value of $s$ 
such that the solutions of
\begin{align}
(s-a)(2\mpi^2-s-a)&=b_s^-(a),\notag\\
(s-a)\bigg(2\mpi^2-s-a+\frac{(s-\mpi^2)^2}{s}\bigg)&=b_s^+(a)\label{schannelsol}
\end{align}
 coincide. This procedure results in
\begin{align}
\label{schannconst}
a&=-41.3\,\mpi^2,\quad s_{\rm max}=27.8\,\mpi^2,\notag\\
b_s^+(a)&=2852\,\mpi^4,\quad b_s^-(a)=1071\,\mpi^4.
\end{align}
$\rho_{su}$ and $\rho_{tu}$ do not yield additional constraints.

\begin{figure}
\centering
\includegraphics[width=\linewidth,clip]{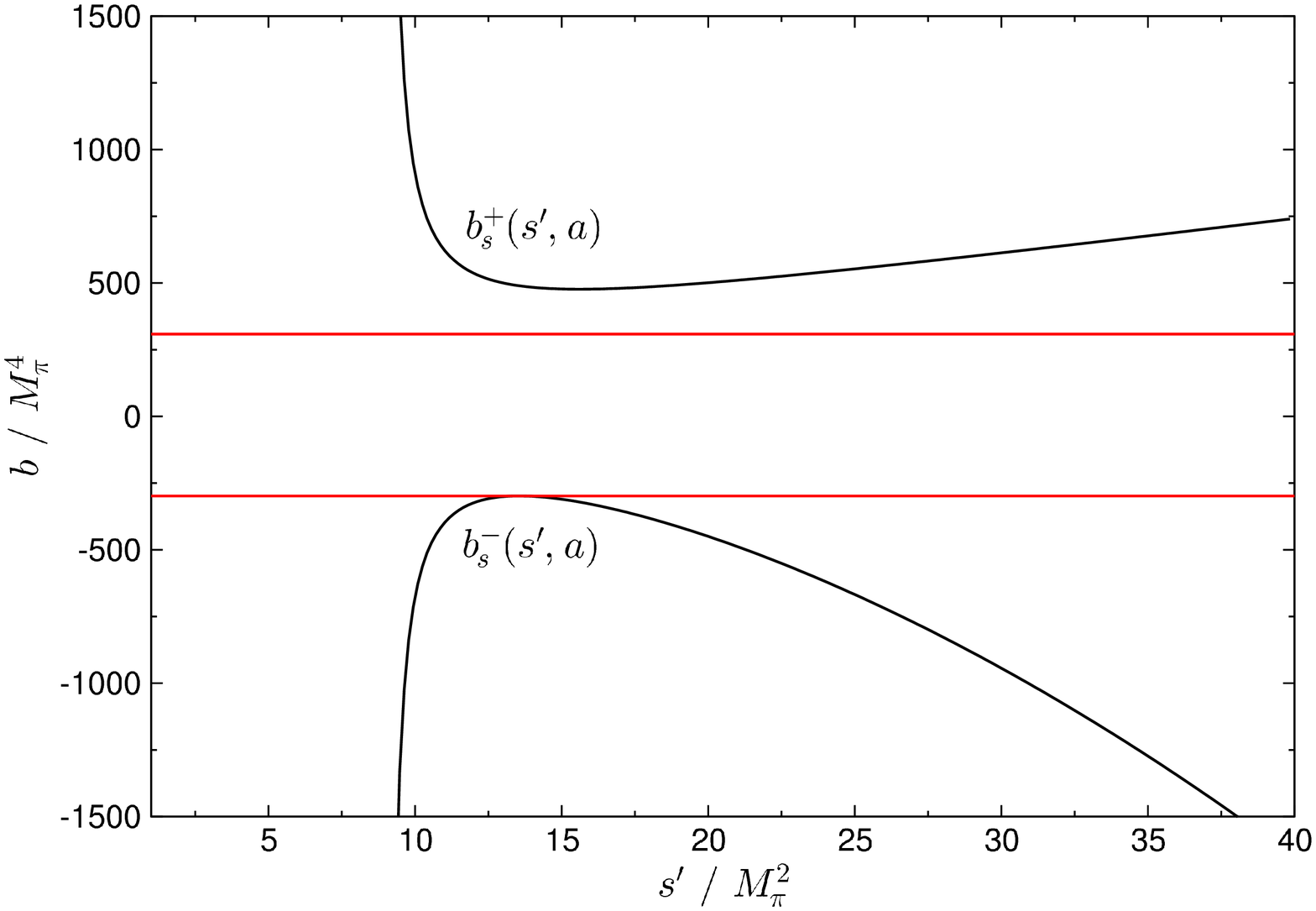}
\includegraphics[width=\linewidth,clip]{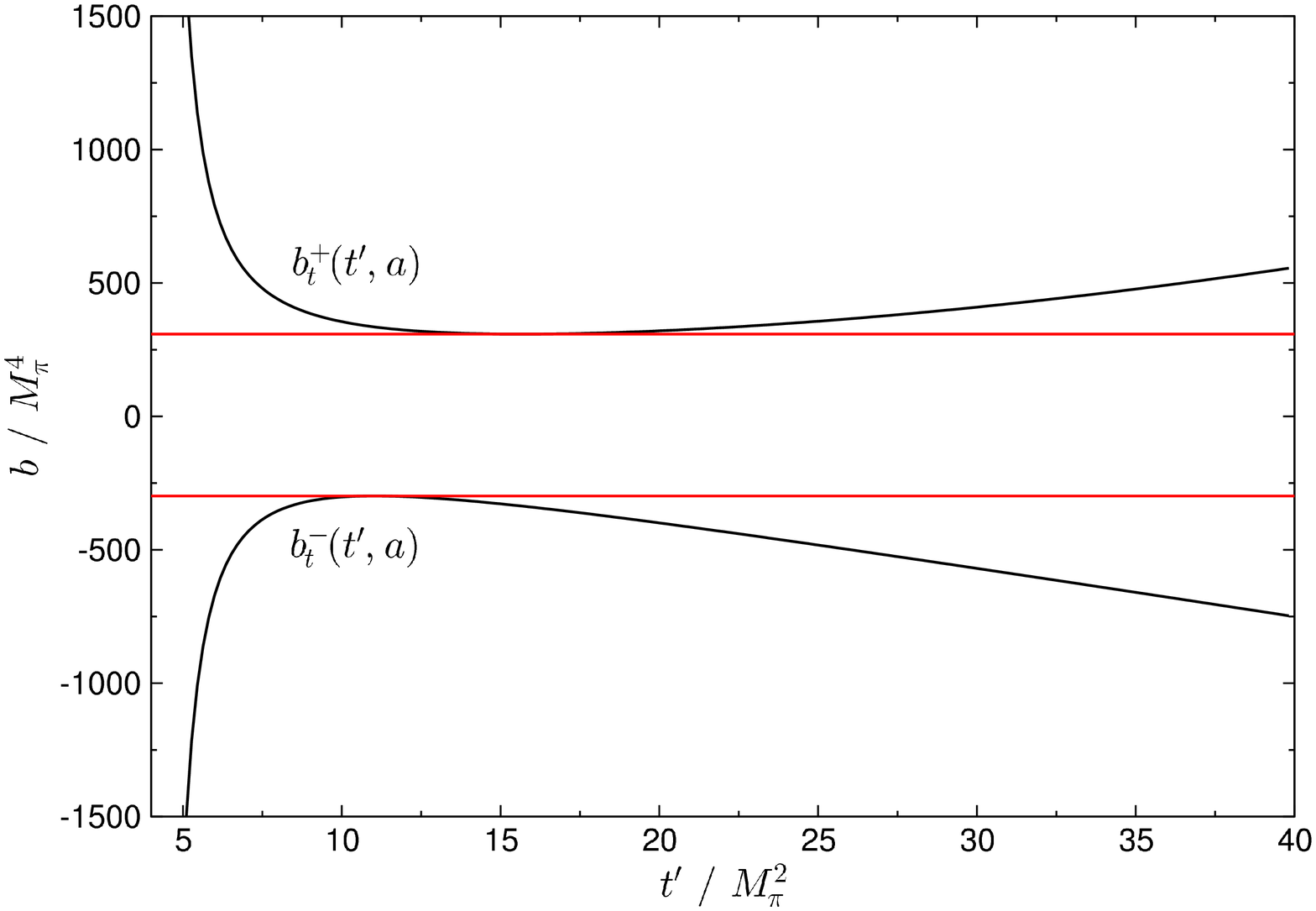}
\caption{Allowed range of $b$ for $a=-9.8\ \mpi^2$ in the $s$-channel projection. The red lines correspond to $b^+_t(a)=308.4\,\mpi^4$ and $b^-_t(a)=-298.1\,\mpi^4$, respectively.}
\label{fig:bplusminus}
\end{figure}

\begin{figure}
\centering
\includegraphics[width=\linewidth,clip]{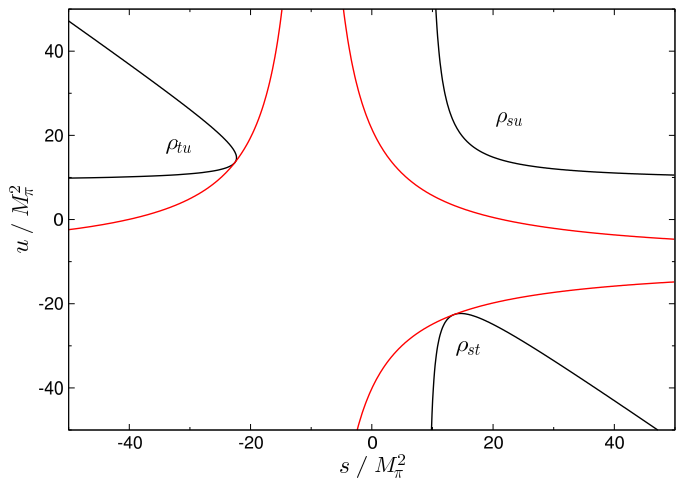}
\caption{Double spectral regions and hyperbolae for $a=-9.8\ \mpi^2$, $b^+_t(a)=308.4\,\mpi^4$ and $b^-_t(a)=-298.1\,\mpi^4$.}
\label{fig:doublespectrals}
\end{figure}

The investigation of the $t$-channel contributions proceeds along the same lines: demanding that the solutions of the two equations
\begin{align}
(s-a)(2\mpi^2-s-a)&=b_t^-(a),\notag\\
(s-a)\bigg(2\mpi^2-s-a+\frac{(s-\mpi^2)^2}{s}\bigg)&=b_t^+(a)
\end{align}
coincide, we find
\begin{align}
a&=-9.8\,\mpi^2,\quad s_{\rm max}=21.4\,\mpi^2,\notag\\
b^+_t(a)&=308.4\,\mpi^4,\quad b^-_t(a)=-298.1\,\mpi^4\label{tchannelres}.
\end{align}
\begin{sloppypar}
Thus, the $s$-channel constraint \eqref{schannconst} is weaker than the $t$-channel restriction \eqref{tchannelres}, as can be deduced from Fig.~\ref{fig:bplusminus}, where the situation for $a=-9.8\,\mpi^2$ is displayed: both $b^+_t$ and $b^-_t$ also lie within the allowed region for the $s$-channel. Since the hyperbolae resulting from the choice \eqref{tchannelres} do not enter the double spectral regions either (see Fig.\ \ref{fig:doublespectrals}), \eqref{tchannelres} constitutes the final answer for the range of validity of the $s$-channel projection.
\end{sloppypar}

\subsection{$\boldsymbol{t}$-channel projection}
\label{subsec:tconvergence}

As we may start from a different set of hyperbolic dispersion relations to perform the $t$-channel projection \eqref{tchanneleq}, there is no need to use the same value of $a$ as in the $s$-channel. To perform the $t$-channel projection, we need to consider
\beq
0\leq z_t^2=\frac{(t-2\mpi^2+2a)^2-4b}{16q_t^2p_t^2}\leq 1.
\eeq
Similarly to the $s$-channel case, we determine the optimal choice of $a$ such that the solutions for $t$  with~$z_t^2=0$ and $z_t^2=1$ are the same. This yields
\begin{align}
a&=-7.5\,\mpi^2,\quad -17.4\,\mpi^2\leq t \leq 51.6\,\mpi^2=1\,{\rm GeV}^2,\notag\\
b^+_t(a)&=298.4\,\mpi^4,\quad b^-_t(a)=-316.8\,\mpi^4.
\label{t_channel_a}
\end{align}
Again, this region is driven by the constraint on the $t$-channel Lehmann ellipse, which provides a stronger constraint than that involving $b_s^+(a)$ and $b_s^-(a)$. We will use the value $a=-7.5 M_\pi^2$ in the rest of the paper.

\begin{figure}
\centering
\includegraphics[width=\linewidth,clip]{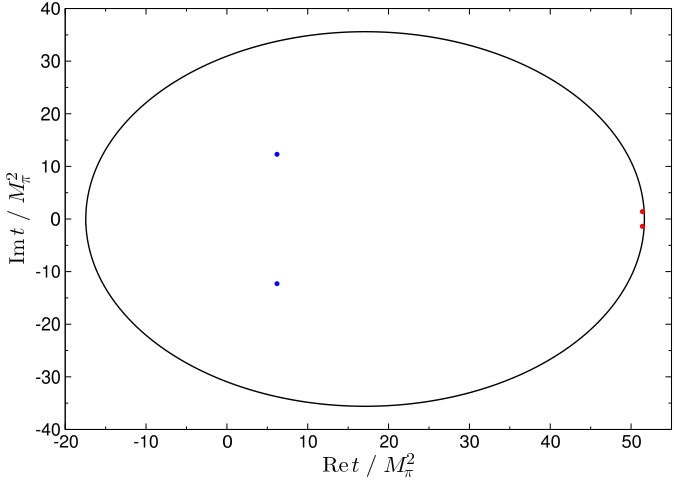}
\caption{Domain of validity in the complex $t$-plane for $a=-7.5\ \mpi^2$. The blue dots refer to the position of the $\sigma$-pole at $t=(6.2\pm i 12.3)\,\mpi^2$ and the red dots to the $f_0$ at $t=(51.4\pm i 1.4)\,\mpi^2$ \cite{CCL}.}
\label{fig:complex_t}
\end{figure}

As we eventually aim to investigate the properties of the $\sigma$, we also need to consider the domain of convergence in the complex plane. We will restrict our analysis to the value of the hyperbola parameter $a=-7.5 M_\pi^2$. Then, the constraints from both $s$-channel and $t$-channel Lehmann ellipses yield ellipses of allowed values for $b$ in the complex $b$-plane. As the ellipse for the $s$-channel contains the ellipse for the $t$-channel, it suffices to consider the latter. 
 The resultant region of the complex $t$-plane is depicted in Fig.~\ref{fig:complex_t}. It safely encompasses the position of the $\sigma$ pole, and marginally that of the $f_0(980)$.  

\begin{sloppypar}
The result that the equations for the $t$-channel are rigorously valid up to $t_{\rm max}=1\,{\rm GeV}^2$ seems to shed doubt on recent dispersive fits \cite{GM_10} to high-statistics $\gamma\gamma\to\pi\pi$ data \cite{Belle07,Belle08} in the energy region up to $1.28\,{\rm GeV}$, where a lot of effort was put into building a good description of $K \bar K$ dynamics above $1\,{\rm GeV}$. One possible way to obtain a higher upper limit on the range of validity of the $\gamma \pi$ Roy equations would be to relax the assumptions about the boundaries of the double spectral functions: if one assumed that the spectral strength of the $2$-pion intermediate states depicted in~Fig.~\ref{fig:boundary} only set in at an effective mass larger than $2\mpi$, the domain of validity would be extended accordingly. For instance, a threshold of ${m_{\rm eff}=3\mpi}$ would produce $t_{\rm max}=82.8\,\mpi^2=(1.27\,{\rm GeV})^2$ at ${a=-12.0\,\mpi^2}$.  
\end{sloppypar}

\section{Muskhelishvili--Omn\`es problem for $\boldsymbol{\gamma\gamma\to\pi\pi}$}
\label{sec:omnes}

We now turn to the resolution of the equations for $\gamma\gamma\to\pi\pi$. We truncate the system at $J=2$ both for $s$- and $t$-channel contributions. The generalization to higher partial waves is straightforward. In this approximation, the equations can be brought into the form
\begin{align}
\label{mo_eq}
 h_{0,+}(t)&=\Delta_{0,+}^{(i)}(t)+\frac{t^{1+i}}{\pi}\int\limits_{4\mpi^2}^\infty\diff t'\frac{\Im h_{0,+}(t')}{t'^{1+i}(t'-t)},\notag\\
 h_{2,+}(t)&=\Delta_{2,+}(t)\notag\\
&+\frac{t^2(t-4\mpi^2)}{\pi}\int\limits_{4\mpi^2}^\infty\diff t'\frac{\Im h_{2,+}(t')}{t'^2(t'-4\mpi^2)(t'-t)},\notag\\
h_{2,-}(t)&=\Delta_{2,-}^{(i)}(t)\\
&+\frac{t^{1+i}(t-4\mpi^2)}{\pi}\int\limits_{4\mpi^2}^\infty\diff t'\frac{\Im h_{2,-}(t')}{t'^{1+i}(t'-4\mpi^2)(t'-t)},\notag
\end{align}
where $i\in\{0,1,2\}$ indicates the number of subtractions, and $\Delta_{J,\pm}(t)$ includes the Born terms, subtraction constants, and integrals involving the imaginary part of other partial waves
\begin{align}
\label{SDwave}
 \Delta_{0,+}^{(i)}(t)&=\tilde N_0^{+\,(i)}(t)+\frac{1}{\pi}\int\limits_{4\mpi^2}^\infty\diff t' \Big(\tilde K_{02}^{++\,(i)}(t,t')\Im h_{2,+}(t')\notag\\
&\qquad+\tilde K_{02}^{+-\,(i)}(t,t')\Im h_{2,-}(t')\Big)\notag\\
&+\frac{1}{\pi}\int\limits_{\mpi^2}^\infty\diff s'\sum\limits_{J'=1,2} \Big(\tilde G_{0J'}^{++\,(i)}(t,s')\Im f_{J',+}(s')\notag\\
&\qquad+\tilde G_{0J'}^{+-\,(i)}(t,s')\Im f_{J',-}(s')\Big),\notag\\
\Delta_{2,+}(t)&=\tilde N_2^{+}(t)+\frac{1}{\pi}\int\limits_{\mpi^2}^\infty\diff s'\hspace{-0.3pt}\sum\limits_{J'=1,2}\hspace{-0.3pt} \Big(\tilde G_{2J'}^{++}(t,s')\Im f_{J',+}(s')\notag\\
&\qquad+\tilde G_{2J'}^{+-}(t,s')\Im f_{J',-}(s')\Big),\notag\\
\Delta_{2,-}^{(i)}(t)&=\tilde N_2^{-\,(i)}(t)\notag\\
&+\frac{1}{\pi}\int\limits_{\mpi^2}^\infty\diff s'\sum\limits_{J'=1,2}\tilde G_{2J'}^{-+\,(i)}(t,s')\Im f_{J',+}(s').
\end{align}
\begin{sloppypar}
Several comments are in order: first, the equation for $h_{2,+}(t)$ is not affected by subtractions. Second, the equations for the $D$-waves decouple, as the corresponding kernels relating them to the $S$-wave or to each other vanish. Conversely, the $S$-wave does not decouple from the $D$-waves: both $D$-waves are needed as input to solve for $h_{0,+}(t)$. The consequence of these observations for the numerical evaluation is that we first solve the equations for the $D$-waves separately, and then we use these solutions to compute $\Delta_{0,+}^{(i)}(t)$
 as input for the $S$-wave.
\end{sloppypar}   

Assuming elastic unitarity
\beq
\label{unitarity}
\Im h_{J,\pm}(t)=\sigma(t) h_{J,\pm}(t)t_J(t)^*
\eeq
with $\pi \pi$ partial waves $t_J(t)$ in the normalization
\beq
t_J(t)=\frac{e^{2i\delta_J(t)}-1}{2i\sigma(t)},
\eeq
the phase of $h_{J,\pm}(t)$ coincides with the $\pi\pi$ phase $\delta_J(t)$ (``Watson's theorem'' \cite{Watson}), and \eqref{mo_eq} takes the form of a Mus\-khelishvili--Omn\`es problem \cite{Muskhelishvili,Omnes}, whose resolution we will review in the following section.

\subsection{Muskhelishvili--Omn\`es problem with finite matching point}
\label{sec:finite}

As we have argued in Sect.~\ref{sec:dom_val}, the domain of validity of \eqref{mo_eq} is restricted to the energy range below $1\,{\rm GeV}$. We will follow here the strategy of \cite{ACGL,piK}, namely to assume that the imaginary parts of the amplitudes are known above a matching point $\tm$, and  solve the equations in the regime between threshold and $\tm$. The input that we will use both for the $s$-channel contributions and the high-energy regime above $\tm$ will be discussed in detail in Sect.~\ref{sec:input}.

\begin{sloppypar}
In this finite-matching-point setup, the solution in terms of Omn\`es functions reads
\begin{align}
\label{omnes_sol}
 h_{0,+}(t)&=\Delta_{0,+}^{(i)}(t)+\frac{t^{1+i}\Omega_0(t)}{\pi}\Bigg\{\int\limits_{\tm}^{\infty}\diff t'\frac{\Im h_{0,+}(t')}{t'^{1+i}(t'-t)|\Omega_{0}(t')|}\notag\\
&+\int\limits_{4\mpi^2}^{\tm}\diff t'\frac{\sin\delta_0(t')\Delta_{0,+}^{(i)}(t')}{t'^{1+i}(t'-t)|\Omega_{0}(t')|}\Bigg\},\notag\\
h_{2,+}(t)&=\Delta_{2,+}(t)+\frac{t^2(t-4\mpi^2)\Omega_2(t)}{\pi}\notag\\
&\times\Bigg\{\int\limits_{4\mpi^2}^{\tm}\diff t'\frac{\sin\delta_2(t')\Delta_{2,+}(t')}{t'^2(t'-4\mpi^2)(t'-t)|\Omega_{2}(t')|}
\notag\\
&+\int\limits_{\tm}^{\infty}\diff t'\frac{\Im h_{2,+}(t')}{t'^2(t'-4\mpi^2)(t'-t)|\Omega_{2}(t')|}\Bigg\},\notag\\
h_{2,-}(t)&=\Delta_{2,-}^{(i)}(t)+\frac{t^{1+i}(t-4\mpi^2)\Omega_2(t)}{\pi}\notag\\
&\times\Bigg\{\int\limits_{4\mpi^2}^{\tm}\diff t'\frac{\sin\delta_2(t')\Delta_{2,-}^{(i)}(t')}{t'^{1+i}(t'-4\mpi^2)(t'-t)|\Omega_{2}(t')|}
\notag\\
&+\int\limits_{\tm}^{\infty}\diff t'\frac{\Im h_{2,-}(t')}{t'^{1+i}(t'-4\mpi^2)(t'-t)|\Omega_{2}(t')|}\Bigg\},
\end{align}
where
\beq
\Omega_J(t)=\exp\Bigg\{\frac{t}{\pi}\int\limits_{4\mpi^2}^{\tm}\diff t'\frac{\delta_J(t')}{t'(t'-t)}\Bigg\}.
\eeq
Defining the physical amplitude by the limit $t\to t+i\eps$, \eqref{omnes_sol} can be rewritten as\footnote{The symbol $\dashint{0pt}$ indicates the principal value of the integral.}
\begin{align}
\label{omnes_sol_mod}
 |h_{0,+}(t)|&=\Delta_{0,+}^{(i)}(t)\cos\delta_0(t)+\frac{t^{1+i}|\Omega_0(t)|}{\pi}\notag\\
&\times\Bigg\{\dashint{7pt}\limits_{4\mpi^2}^{\tm}\diff t'\frac{\sin\delta_0(t')\Delta_{0,+}^{(i)}(t')}{t'^{1+i}(t'-t)|\Omega_{0}(t')|}\notag\\
&+\int\limits_{\tm}^{\infty}\diff t'\frac{\Im h_{0,+}(t')}{t'^{1+i}(t'-t)|\Omega_{0}(t')|}\Bigg\},
\end{align}
and similarly for the $D$-waves. 
There are several subtleties related to the behavior of $|\Omega_J(t)|$ for $t\to \tm$ \cite{piK}
\beq
|\Omega_J(t)|\sim |\tm-t|^{\frac{\delta_J(\tm)}{\pi}},
\eeq
in particular the integrals in \eqref{omnes_sol_mod} diverge for $\delta_J(\tm)>\pi$. In such a case, there are non-trivial homogeneous solutions whose coefficients can be used to absorb these divergences, but, of course, the presence of such solutions introduces undetermined constants in our result for $\Omega_J(t)$. For instance, the solution for $h_{0,+}(t)$ for $\pi<\delta_0(\tm)<2\pi$ involves one free parameter $\alpha$,
\begin{align}
 |h_{0,+}(t)|&=\Delta_{0,+}^{(i)}(t)\cos\delta_0(t)+\frac{t^{1+i}|\Omega_0(t)|}{(\tm-t)\pi}\notag\\
&\times\Bigg\{\alpha+t\dashint{7pt}\limits_{4\mpi^2}^{\tm}\diff t'\frac{(\tm-t')\sin\delta_0(t')\Delta_{0,+}^{(i)}(t')}{t'^{2+i}(t'-t)|\Omega_{0}(t')|}\notag\\
&+t\int\limits_{\tm}^{\infty}\diff t'\frac{(\tm-t')\Im h_{0,+}(t')}{t'^{2+i}(t'-t)|\Omega_{0}(t')|}\Bigg\}.
\end{align} 
$\alpha$ can be fixed if one assumes, in addition, knowledge of the derivative of $h_{0,+}(t)$ at $\tm$. Conversely, the fact that the phases of the $I=2$ partial waves are negative induces different complications, which we will address in Sect.~\ref{sec:I2_sumrule}.
\end{sloppypar}

The part of the integrals in \eqref{omnes_sol} involving the pion polarizabilities can be explicitly performed based on the spectral representation of the Omn\`es functions
\begin{align}
\label{omnes_spectral}
\Omega_J^{-1}(t)&=-\frac{1}{\pi}\int\limits_{4\mpi^2}^{\tm} \diff t'\frac{\sin\delta_J(t')}{(t'-t)|\Omega_J(t')|},\notag\\
\Omega_J^{-1}(t)&=1-\frac{t}{\pi}\int\limits_{4\mpi^2}^{\tm} \diff t'\frac{\sin\delta_J(t')}{t'(t'-t)|\Omega_J(t')|},\\
\Omega_J^{-1}(t)&=1-t\,\dot\Omega_J(0)-\frac{t^2}{\pi}\int\limits_{4\mpi^2}^{\tm} \diff t'\frac{\sin\delta_J(t')}{t'^2(t'-t)|\Omega_J(t')|},\notag
\end{align}
where the dot denotes the derivative with respect to $t$. The results of this modification, which,  in particular, only involve integrals over
\beq
\label{Delta_tilde}
\tilde\Delta_{J,\pm}(t)=\Delta_{J,\pm}(t)-\Delta \tilde N_J^{\pm}(t),
\eeq 
are summarized in App.~\ref{App:omnes_sol_I0} (with $\Delta \tilde N_J^{\pm}(t)$ defined in App.~\ref{app:kernel_tchannel}).

\subsection{Sum rules for $\boldsymbol{I=2}$}
\label{sec:I2_sumrule}

If $\delta_J(\tm)<0$, $\Omega_0(t)$ diverges at $\tm$, such that \eqref{omnes_sol} breaks down. However,
we may rewrite the solution as
\begin{align}
h_{0,+}(t)&=\Delta_{0,+}^{(i)}(t)+\frac{t^{1+i}\Omega_0(t)(\tm-t)}{\pi}\notag\\
&\times\Bigg\{\int\limits_{4\mpi^2}^{\tm}\diff t'\frac{\sin\delta_0(t')\Delta_{0,+}^{(i)}(t')}{t'^{1+i}(\tm-t')(t'-t)|\Omega_{0}(t')|}\notag\\
&+\int\limits_{\tm}^{\infty}\diff t'\frac{\Im h_{0,+}(t')}{t'^{1+i}(\tm-t')(t'-t)|\Omega_{0}(t')|}\Bigg\}\notag\\
&+\tilde\alpha^{(i)}\left(\frac{t}{\tm}\right)^{1+i}\Omega_0(t),
\end{align}
where
\begin{align}
\tilde\alpha^{(i)}&=\frac{\tm^{1+i}}{\pi}\int\limits_{4\mpi^2}^{\tm}\diff t'\frac{\sin\delta_0(t')\Delta_{0,+}^{(i)}(t')}{t'^{1+i}(t'-\tm)|\Omega_{0}(t')|}\notag\\
&+\frac{\tm^{1+i}}{\pi}\int\limits_{\tm}^{\infty}\diff t'\frac{\Im h_{0,+}(t')}{t'^{1+i}(t'-\tm)|\Omega_{0}(t')|}.
\end{align}
Demanding continuity at the matching point implies that $\tilde \alpha^{(i)}=0$, since otherwise $h_{0,+}(t)$ would diverge at $\tm$. Indeed, multiplying the first equation of \eqref{omnes_sol} with $\Omega_0^{-1}(t)$ and evaluating it at $t=\tm$ explicitly proves that $\tilde \alpha^{(i)}=0$ must hold in order for $h_{0,+}(\tm)$ to be finite. The final results for all amplitudes as well as the explicit form of all of the possible sum rules which could be obtained are given in App.~\ref{App:omnes_sol_I2}.

One case that is of particular interest is the $I=2$ $S$-waves, where the $\pi \pi$ phase shift is substantial, and negative, at the matching point. Thus, we obtain from the once- and twice-subtracted versions of the Muskhelishvili--Omn\`es representation
\begin{align}
\label{sumrule1}
 0&=\frac{\mpi}{2\alpha}\tm(\alpha_1-\beta_1)^{I=2}+I^{(1)},\notag\\
0&=\frac{\mpi}{2\alpha}\tm(1-\tm\,\dot \Omega_0(0))(\alpha_1-\beta_1)^{I=2}\notag\\
&+\frac{\mpi}{24\alpha}\tm^2(\alpha_2-\beta_2)^{I=2} +I^{(2)},
\end{align}
where
\begin{align}
 I^{(i)}&=\frac{\tm^{1+i}}{\pi}\Bigg\{\int\limits_{4\mpi^2}^{\tm}\diff t'\frac{\sin\delta_0(t')\tilde\Delta_{0,+}^{(i)}(t')}{t'^{1+i}(t'-\tm)|\Omega_{0}(t')|}\notag\\
&+\int\limits_{\tm}^{\infty}\diff t'\frac{\Im h_{0,+}(t')}{t'^{1+i}(t'-\tm)|\Omega_{0}(t')|}\Bigg\}, \quad i\in\{1,2\}.
\end{align}
As we shall see below, the second sum rule, in particular, provides a useful and novel constraint on $(\alpha_2 - \beta_2)^{I=2}$. 

\subsection{Uniqueness and comparison to $\boldsymbol{\pi\pi}$ Roy equations} 

\begin{sloppypar}
The pattern in which free parameters emerge in the Omn\`es solutions is reminiscent of the results concerning the uniqueness properties of the solutions of $\pi\pi$ Roy equations presented in \cite{EpeleWanders,GasserWanders,Wanders}. There, an additional free parameter occurs each time the phase at the matching point crosses an integer multiple of $\pi/2$. Indeed, to derive this result the Roy equations are linearized, and in the one-channel approximation one finds that the difference
\end{sloppypar}
\beq
\phi(s)=\frac{\delta'(s)-\delta(s)}{\sigma(s)}
\eeq
between two solutions $\delta(s)$ and $\delta'(s)$ for the $\pi\pi$ phase shift must fulfill \cite{GasserWanders}
\beq
\cos 2\delta(s)\phi(s)=\frac{s-4\mpi^2}{\pi}\dashint{7pt}\limits_{4\mpi^2}^{\sm}\diff s'\frac{\sin 2\delta(s')\phi(s')}{(s'-4\mpi^2)(s'-s)},
\eeq
where $\sm$ is the matching point. In the $\pi\pi$ case the phase shift is the quantity that is to be determined, while in $\gamma\gamma\to\pi\pi$ that phase is input and the modulus of the amplitude $|h(t)|$ unknown, but, as we shall now see, the mathematical structure is the same up to a factor of $2$. 

As an example we consider the unsubtracted equation for the $S$-wave. In $\gamma\gamma\to\pi\pi$ the difference
\beq
\psi_{0,+}(t)= |h'_{0,+}(t)|-|h_{0,+}(t)|
\eeq
between two solutions of \eqref{mo_eq} obeys
\beq
\cos\delta_{0}(t)\psi_{0,+}(t)=\frac{t}{\pi}\dashint{7pt}\limits_{4\mpi^2}^{\tm}\diff t'\frac{\sin\delta_0(t')\psi_{0,+}(t')}{t'(t'-t)}.
\eeq
The presence of $\sin\delta_0(t')$, rather than the $\sin 2 \delta(s')$ of the $\pi \pi$ case,
explains why the multiplicity of solutions in $\pi\pi$ scattering is $\lfloor 2\delta(\sm)/\pi \rfloor$ instead of $\lfloor \delta_0(\tm)/\pi \rfloor$ in $\gamma\gamma\to\pi\pi$, where $\lfloor x \rfloor$ denotes the largest integer $\leq x$.

In addition, there is also an analog of the sum rules discussed in Sect.~\ref{sec:I2_sumrule} in $\pi\pi$ Roy equations. In the single-channel approximation $\phi(s)$ vanishes if $\delta(\sm)<\pi/2$. If the effects of the coupling to other partial waves are taken into account, the phase-shift difference for a channel $i$ with $\delta_i(\sm)<\pi/2$ may be written as \cite{Wanders}
\begin{align}
\phi_i(s)&=(s-4\mpi^2)G_i(s)H_i(s),\notag\\
G_i(s)&=\exp\Bigg\{\frac{2}{\pi}\int\limits_{4\mpi^2}^{\sm}\diff s'\frac{\delta_i(s')}{s'-s}\Bigg\},
\end{align}
where $H_i(s)$ includes information on other partial waves. Once again, $G_i(\sm)$ diverges if $\delta_i(\sm)<0$, such that 
\beq
H_i(\sm)=0
\eeq
in this case. The difference as compared to $\gamma\gamma\to\pi\pi$ is that in $\pi \pi$ the Omn\`es representation is only available for phase-shift differences and not for the phase shifts themselves. Thus, the constraint manifests itself rather subtly by reducing the number of free parameters in the manifold of solutions. However, the mathematical input that leads to this constraint, namely continuity of the Muskhelishvili--Omn\`es representation at the matching point, is the same.

\subsection{Comparison to previous work}
\label{sec:comparison}

The key result of our derivation of dispersion relations for $\gamma \pi$ scattering and $\gamma \gamma \rightarrow \pi \pi$ based on Roy--Steiner equations is that there is a term that couples $S$- and $D$-waves, cf.~\eqref{SDwave}. In all previous dispersive treatments of $\gamma \gamma \rightarrow \pi \pi$  each partial wave was considered separately. Moreover, to the best of our knowledge, this is the first time that a finite-matching-point representation has been employed for $\gamma\gamma\to\pi\pi$. The practical consequences of both these developments will be discussed in Sects.~\ref{sec:input} and~\ref{sec:results}. However, we can consider the limit $\tm\to\infty$ in our equations in order to delineate the differences between our formalism and the recent works \cite{Pennington06,ORS_08,GM_10}, which also used dispersion relations to attack the problem $\gamma \gamma \to \pi \pi$.

In \cite{Pennington06}, a once-subtracted\footnote{Note that  in this work we do not count powers of $t'$ or $t'-4\mpi^2$ that are present for kinematical reasons alone and thus do not require any subtraction constants. This is not always the convention employed in the literature.} dispersion relation for the $S$-wave is considered. The subtraction constant is fixed by assuming $h_{0,+}(t)\propto t_0(t)$ and using ChPT information on the Adler zero of the $\pi\pi$ amplitude. This representation depends quite strongly on the details of the $\pi\pi$ phase above the $K\bar K$ threshold already at energies $\gtrsim 0.5\,{\rm GeV}$. For this reason, another subtraction was performed in \cite{ORS_08} at the energy $t_1$ where the $\pi\pi$ phase crosses $\pi$, the subtraction constant being fixed by the requirement that the cross section at $t_1$ does not become outrageously large. The Born terms as well as vector and axial-vector resonances were used to approximate the left-hand cut.

In \cite{GM_10}, an Omn\`es representation for $S$- and $D$-waves is constructed that explicitly takes into account the $K\bar K$ channel, and in addition includes tensor resonances in the description of the left-hand cut. For the $S$-waves two subtractions are performed, while the $D$-waves are treated differently for the two isospin channels: for $I=2$, no subtraction constants for $h_{2,+}$ and $h_{2,-}$ are provided (in our conventions this corresponds to the unsubtracted case for $h_{2,-}$), while for $I=0$ an additional subtraction in both partial waves is performed. Thus the treatment of $I=0$ corresponds to our once-subtracted case for $h_{2,-}$. But the Roy--Steiner analysis shows that the subtraction constant for $h_{2,+}$ cannot be related to dipole and quadrupole polarizabilities, as the equations for this partial wave are not affected by the corresponding subtractions. For this reason, the subtraction constants in \cite{GM_10} determined by fits to data (together with several chiral constraints) can be translated into pion polarizabilities, but in general the converse is not possible. We have checked explicitly that our results agree with \cite{GM_10} in the limit $\tm\to\infty$, once the $K\bar K$ channel is switched off and the additional subtraction in $h_{2,+}$ dropped.

\section{Photon coupling of the $\boldsymbol{\sigma}$ resonance}
\label{sec:sigma}

We define the $\sigma \pi\pi$ coupling constant $g_{\sigma\pi\pi}$ such that the full isospin $I=0$ $\pi\pi$ scattering amplitude on the second Riemann sheet $T^0_{\rm II}$ near the position of the $\sigma$ pole 
\beq
t_\sigma=\bigg(M_\sigma-i\frac{\Gamma_\sigma}{2}\bigg)^2
\eeq
can be written as
\beq
T^0_{\rm II}=32\pi\sum\limits_{J=0}^\infty(2J+1)t_{J,{\rm II}}^0(t)P_J(z_t)=\frac{g_{\sigma\pi\pi}^2}{t_\sigma-t}.
\eeq 
As the $\sigma$-pole occurs in the $S$-wave, all other partial waves only contribute to the background. Similarly, in $\gamma\gamma\to\pi\pi$ we take
\beq
e^2H^0_{++,{\rm II}}=\frac{e^2g_{\sigma\pi\pi}g_{\sigma\gamma\gamma}}{t_\sigma-t}.
\eeq
In these conventions, the widths in the narrow-width approximation are
\beq
\label{widths}
\Gamma_{\sigma\pi\pi}=\frac{|g_{\sigma\pi\pi}|^2}{32\pi M_\sigma}\sqrt{1-\frac{4\mpi^2}{M_\sigma^2}},\quad \Gamma_{\sigma\gamma\gamma}=\frac{\pi\alpha^2|g_{\sigma\gamma\gamma}|^2}{M_\sigma}.
\eeq
Note that since the large strong width of the $\sigma$ renders the applicability of these relations questionable, we use $g_{\sigma\pi\pi}$ as determined from the residue of the pole. Although the formula for $\Gamma_{\sigma\gamma\gamma}$ and $g_{\sigma\gamma\gamma}$ suffers from the same deficiency, it is conventionally employed in the literature to illustrate the relation between the two quantities. As the direct determination of $\Gamma_{\sigma\gamma\gamma}$ from the position of the pole is not possible in view of the large strong width, we follow this convention. The coupling constant itself can always be recovered by means of \eqref{widths}.

The analytic continuation of $h_{0,+}^0(t)$ into the complex plane is given by~\cite{ORS_08}\footnote{We neglect $\gamma\gamma$ intermediate states in the unitarity relation. This is the same approximation as used in \cite{CCL}, where the $\sigma$ pole is deduced from $\pi\pi$ scattering with electromagnetic interactions switched off. Since the width corresponding to the $\gamma\gamma$ channel amounts only to a few keV, the ensuing shift of the pole is much smaller than the uncertainty of its position as quoted in \cite{CCL}.}
\beq
h_{0,+,{\rm II}}^0(t)=(1-2i\sigma(t)t_{0,{\rm II}}^0(t))h_{0,+,{\rm I}}^0(t),
\eeq
such that
\beq
\frac{g_{\sigma\gamma\gamma}^2}{g_{\sigma\pi\pi}^2}=-\left(\frac{\sigma(t_\sigma)}{16\pi}\right)^2(h^0_{0,+}(t_\sigma))^2,
\eeq
where $h^0_{0,+}$ is evaluated on the first Riemann sheet. Assuming the position of the $\sigma$ pole and its coupling constant to two pions is known, we can thus infer $g_{\sigma\gamma\gamma}$ (and hence $\Gamma_{\sigma\gamma\gamma}$) from the value of the $I=0$ $S$-wave of $\gamma\gamma\to\pi\pi$ evaluated at $t_\sigma$.
 
\section{Input}
\label{sec:input}

To solve the Roy--Steiner equations for the $\gamma\gamma\to\pi\pi$ partial waves we must specify the input for $\Im f_{J,\pm}(s)$ in the whole energy range and for $\Im h_{J,\pm}(t)$ above the matching point.  One could, in the spirit of the Roy--Steiner analysis for $\pi K$ scattering \cite{piK},  consider the equations for $s$- and $t$-channel partial waves simultaneously, and determine a solution of the whole system by an iterative procedure. However, the Roy--Steiner equations for $\gamma\pi\to\gamma\pi$ are less powerful than those for $\pi K\to \pi K$, since the only contributions that can be obtained in the $s$-channel without additional input are determined by elastic unitarity. In the $\gamma \pi$ case, $\gamma\pi$ intermediate states are suppressed by $e^2$ and thus expected to be numerically negligible. By comparison, in the $\pi K$ case, $\pi K$ intermediate states dominate the unitarity relation at least up to $1\,{\rm GeV}$ \cite{piK}. 

For this reason, we will drop $\gamma\pi$ intermediate states altogether and content ourselves with the resonance description of the imaginary parts of the Compton-scattering amplitudes constructed in \cite{GM_10}, where the resonance contributions are eventually included in a spectral representation with an integration cutoff of $-5\,{\rm GeV}^2$. In our framework the effect of the resonance description of  $\Im f_{J,\pm}(s)$ on $h_{J,\pm}(t)$ can be directly read off from \eqref{tchanneleq} or \eqref{SDwave}, cf.~Sect.~\ref{sec:res}. Physically, one can understand this summation of resonances as an effective description of multi-pion states in the $s$-channel for $\gamma\pi\to\gamma\pi$, or, equivalently, in the $t$-channel for $\gamma\gamma\to\pi\pi$, amounting to approximating the multi-pion cuts by a sum of poles. This model could be improved upon at low energies by explicitly incorporating the $2$- and $3$-pion intermediate states, and using ChPT to constrain their interactions. However, these intermediate states enter the $\gamma \pi$ amplitude at $\Order(e^2p^6)$ and $\Order(e^2p^4)$ in the chiral counting, respectively, such that we will leave their incorporation to future work.      

We now turn to the input for $\Im h_{J,\pm}(t)$. We choose the matching point as
\beq
\sqrt{\tm}=0.98\,{\rm GeV},
\eeq
which, on the one hand, ensures that $\delta_0(\tm)<\pi$, avoiding a free parameter in the Omn\`es solution, and, on the other hand, extends the energy range as far as possible. As the cross section above $1\,{\rm GeV}$ is dominated by the $f_2(1270)$ resonance, we put $h_{J,\pm}(t)=0$ above $\tm$ for all partial waves except for $h_{2,-}^{I=0}(t)$, which we match to a Breit--Wigner description of the $f_2(1270)$, cf.~Sect.~\ref{sec:1270}. As we will show in Sect.~\ref{sec:results}, this approximation allows for a reasonable description of the cross section. A more detailed description of this region is not necessary, because our subtracted dispersion relations emphasize the low-energy region which is the focus of this work.  

As input for the $\pi\pi$ phases we use the results of an extended Roy-equation analysis of $\pi\pi$ scattering \cite{Bern}, which in particular ensures that the phases and the pole position of the $\sigma$ are consistent, since \cite{Bern} coincides perfectly with the older analysis \cite{CGL} at low energies. (The impact of the high-energy region in $\pi\pi$ scattering on the $\sigma$-pole position was shown to be negligible in~\cite{CCL}.) To estimate the uncertainties due to the $\pi\pi$ input, we also consider the $\pi\pi$ phases determined in a recent study of Roy-like equations \cite{Madrid}. The parameters of the $\sigma$ resonances corresponding to both approaches, which we will refer to as CCL and GKPRY, respectively, are given in Table~\ref{table:sigma}, and are consistent within error bars.

\begin{table}
\centering
\begin{tabular}{cccc}
\hline\hline
& $M_\sigma$ & $\Gamma_\sigma$ & $g_{\sigma\pi\pi}/\sqrt{2}$ \\\hline
CCL\hspace{1.3pt}\cite{CCL,Cap} & $441^{+16}_{-8}\,{\rm MeV}$ & $544^{+18}_{-25}\,{\rm MeV}$ & $3.3\,{\rm GeV}$\\
GKPRY\hspace{1.3pt}\cite{EGKP10} & $457^{+14}_{-13}\,{\rm MeV}$ & $558^{+22}_{-14}\,{\rm MeV}$ & $3.59^{+0.11}_{-0.13}\,{\rm GeV}$\\
\hline\hline
\end{tabular}
\caption{Mass, width, and $\pi\pi$ coupling constant of the $\sigma$.}
\label{table:sigma}
\end{table}

Finally, our results depend on the input chosen for the pion polarizabilities. For definiteness, we will consider the two sets of parameters compiled in Table~\ref{table:pol}, which we refer to as ChPT and GMM, respectively. The polarizabilities in the isospin basis follow from \eqref{isospin}. Note that in \cite{GM_10} the dipole polarizability of the charged pion was only allowed to vary within the range of the ChPT prediction. Unfortunately, the charged-pion quadrupole polarizability $\alpha_2-\beta_2$ is rather sensitive to low-energy constants (LECs): the first number, $16.2 \cdot 10^{-4}\,{\rm fm}^5$, corresponds to the resonance-saturation model of \cite{GIS06}, while taking the LECs from \cite{Bijnens96} yields $21.6 \cdot 10^{-4}\,{\rm fm}^5$. 

\begin{table}
\centering
\begin{tabular}{ccc}
\hline\hline
& ChPT~\cite{GIS05,GIS06} & GMM~\cite{GM_10} \\\hline
$(\alpha_1-\beta_1)^{\pi^0}$ & $-1.9\pm 0.2$ & $-1.25\pm 0.17$\\
$(\alpha_1+\beta_1)^{\pi^0}$ & $1.1\pm 0.3$ & $1.22\pm 0.12$\\
$(\alpha_2-\beta_2)^{\pi^0}$ & $37.6\pm 3.3$ & $32.1\pm 2.1$\\
$(\alpha_2+\beta_2)^{\pi^0}$ & $0.037\pm 0.003$ & $-0.19\pm 0.02$\\\hline
$(\alpha_1-\beta_1)^{\pi^{\pm}}$ & $5.7\pm 1.0$ & $4.7$\\
$(\alpha_1+\beta_1)^{\pi^{\pm}}$ & $0.16\, [0.16]$ & $0.19\pm 0.09$\\
$(\alpha_2-\beta_2)^{\pi^{\pm}}$ & $16.2\, [21.6]$ & $14.7\pm 2.1$\\
$(\alpha_2+\beta_2)^{\pi^{\pm}}$ & $-0.001\, [-0.001]$ & $0.11\pm 0.03$\\
\hline\hline
\end{tabular}
\caption{Dipole and quadrupole pion polarizabilities in units of $10^{-4}{\rm fm}^3$ and $10^{-4}{\rm fm}^5$, respectively. The numbers in brackets refer to the LECs from \cite{Bijnens96}.}
\label{table:pol}
\end{table}

\subsection{Resonances in $\boldsymbol{\gamma\pi\to\gamma\pi}$}
\label{sec:res}

We use the resonance model constructed in \cite{GM_10}, with the contribution of vector (V), axial-vector (A), tensor (T), and axial-tensor ($\rm T_A$) resonances to the Compton-scattering partial waves in the narrow-width approximation, to define the imaginary part of the $\gamma\pi\to\gamma\pi$ amplitudes
\begin{align}
\label{fJ_res}
 \Im f^{\rm V}_{J,\pm}(s)&=\pm \frac{2}{3}\pi C_{\rm V}(\mv^2-\mpi^2)^2\delta(s-\mv^2)\delta_{J1},\notag\\
 \Im f^{\rm A}_{J,\pm}(s)&=\frac{2}{3}\pi C_{\rm A}(\ma^2-\mpi^2)^2\delta(s-\ma^2)\delta_{J1},\notag\\
 \Im f^{\rm T}_{J,\pm}(s)&=\pm \frac{2}{5}\pi C_{\rm T}\frac{(\mt^2-\mpi^2)^4}{\mt^2}\delta(s-\mt^2)\delta_{J2},\notag\\
 \Im f^{\rm T_A}_{J,\pm}(s)&= \frac{2}{5}\pi C_{\rm T_A}\frac{(\mta^2-\mpi^2)^4}{\mta^2}\delta(s-\mta^2)\delta_{J2},
\end{align}
where $m_i$, $i\in\{{\rm V}, {\rm A}, {\rm T}, {\rm T_A}\}$, denotes the mass of the resonance, and the coupling constants $C_i$ are related to the widths $\Gamma_i$ by
\begin{align}
\Gamma_{\rm V}&=\alpha C_{\rm V}\frac{(\mv^2-\mpi^2)^3}{3\mv^3},\quad  \Gamma_{\rm A}=\alpha C_{\rm A}\frac{(\ma^2-\mpi^2)^3}{3\ma^3},\notag\\
\Gamma_{\rm T}&=\alpha C_{\rm T}\frac{(\mt^2-\mpi^2)^5}{5\mt^5},\quad  \Gamma_{\rm T_A}=\alpha C_{\rm T_A}\frac{(\mta^2-\mpi^2)^5}{5\mta^5}.
\end{align}
Inserting \eqref{fJ_res} into \eqref{tchanneleq} yields
\begin{align}
\label{hJ_res}
h^{\rm V}_{J,+}(t)&=\frac{2}{3}C_{\rm V}(\mv^2-\mpi^2)^2\Big(\hspace{-0.7pt}\tilde G_{J1}^{++}(t,\mv^2)-\tilde G_{J1}^{+-}(t,\mv^2)\hspace{-0.7pt}\Big),\notag\\
h^{\rm V}_{J,-}(t)&=\frac{2}{3}C_{\rm V}(\mv^2-\mpi^2)^2\tilde G_{J1}^{-+}(t,\mv^2),\notag\\
h^{\rm A}_{J,+}(t)&=\frac{2}{3}C_{\rm A}(\ma^2-\mpi^2)^2\Big(\hspace{-0.7pt}\tilde G_{J1}^{++}(t,\ma^2)+\tilde G_{J1}^{+-}(t,\ma^2)\hspace{-0.7pt}\Big),\notag\\
h^{\rm A}_{J,-}(t)&=\frac{2}{3}C_{\rm A}(\ma^2-\mpi^2)^2\tilde G_{J1}^{-+}(t,\ma^2),\notag\\
h^{\rm T}_{J,+}(t)&=\frac{2}{5}C_{\rm T}\frac{(\mt^2-\mpi^2)^4}{\mt^2}\Big(\hspace{-0.7pt}\tilde G_{J2}^{++}(t,\mt^2)-\tilde G_{J2}^{+-}(t,\mt^2)\hspace{-0.7pt}\Big),\notag\\
h^{\rm T}_{J,-}(t)&=\frac{2}{5}C_{\rm T}\frac{(\mt^2-\mpi^2)^4}{\mt^2}\tilde G_{J2}^{-+}(t,\mt^2),\notag\\
h^{\rm T_A}_{J,+}(t)&=\frac{2}{5}C_{\rm T_A}\frac{(\mta^2-\mpi^2)^4}{\mta^2}\notag\\
&\times\Big(\hspace{-0.7pt}\tilde G_{J2}^{++}(t,\mta^2)+\tilde G_{J2}^{+-}(t,\mta^2)\hspace{-0.7pt}\Big),\notag\\
h^{\rm T_A}_{J,-}(t)&=\frac{2}{5}C_{\rm T_A}\frac{(\mta^2-\mpi^2)^4}{\mta^2}\tilde G_{J2}^{-+}(t,\mta^2).
\end{align}
We include all resonances listed in \cite{GM_10}. We have checked that \eqref{hJ_res} agrees with the results quoted in \cite{GM_10}: once the ambiguous term linear in $t$ in $h_{0,+}(t)$ in \cite{GM_10} is removed, we recover that result by taking the limit $a\to\infty$ of our unsubtracted kernel functions.

\subsection{Including the $\boldsymbol{f_2(1270)}$}
\label{sec:1270}

To incorporate the $D$-wave resonance $f_2(1270)$ we follow \cite{Drechsel:1999rf}. Starting from 
\beq
\Lagr_{\rm TPP}=C_{\rm T}^\pi T^{\mu\nu}\partial_\mu P\partial_\nu P,\quad \Lagr_{\text{T}\gamma\gamma}=e^2C_{\rm T}^\gamma T^{\mu\nu}F_{\mu\alpha}F_\nu^{\,\,\,\alpha}
\eeq
to describe the coupling of a tensor resonance to pseudoscalars and photons, respectively, we find
\begin{align}
A&=-\frac{C_{\rm T}^\pi C_{\rm T}^\gamma}{6(t-\mt^2)}\bigg\{4\mpi^2\bigg(4-\frac{t}{\mt^2}\bigg)-t\bigg(5-\frac{2t^2}{\mt^4}\bigg)\bigg\},\notag\\
B&=\frac{C_{\rm T}^\pi C_{\rm T}^\gamma}{4(t-\mt^2)},
\end{align}
such that
\begin{align}
\label{f2_offshell}
H_{++}&=-\frac{C_{\rm T}^\pi C_{\rm T}^\gamma t}{6\mt^4}\Big(t(t+\mt^2)-2\mt^2\mpi^2\Big),\notag\\ 
H_{+-}&=\frac{C_{\rm T}^\pi C_{\rm T}^\gamma}{4}\frac{t^2\sigma(t)^2}{t-\mt^2}(1-z_t^2).
\end{align}
We see that a resonant contribution only occurs in $h_{2,-}(t)$, while the non-resonant background in $h_{0,+}(t)$ can be discarded. Taking the full width of the $f_2(1270)$ into account and dropping the non-resonant background, we obtain
\beq
\label{f2model}
h_{2,-}^{f_2}(t)=\frac{C_{f_2}^\pi C_{f_2}^\gamma}{5\sqrt{6}}\frac{m_{f_2}^4\sigma(m_{f_2}^2)^2}{t-m_{f_2}^2+i m_{f_2}\Gamma_{f_2}}.
\eeq
In fact, in Sect.~\ref{sec:total_cross_section} we will restore the background in order to describe the cross section for ${\gamma\gamma\to\pi^+\pi^-}$ above the matching point.
The coupling constants can be determined from the partial widths
\begin{align}
\label{width}
\Gamma_{f_2\to\pi\pi}&=\frac{\big(C_{f_2}^\pi\big)^2}{960\pi}\frac{\big(m_{f_2}^2-4\mpi^2\big)^{\frac{5}{2}}}{m_{f_2}^2},\notag\\ 
\Gamma_{f_2\to\gamma\gamma}&=\frac{\pi}{5}\alpha^2\big(C_{f_2}^\gamma\big)^2m_{f_2}^3.
\end{align}
For the $f_2$ parameters we use as input \cite{PDG10}
\begin{align}
m_{f_2}&=1275.1\,{\rm MeV},\quad \Gamma_{f_2}=185.1\,{\rm MeV},\notag\\
\Gamma_{f_2\to\pi\pi}&=156.9\,{\rm MeV},\quad \Gamma_{f_2\to\gamma\gamma}=3.03\,{\rm keV},
\end{align}
such that
\beq
|C_{f_2}^\pi|=16.06\,{\rm GeV}^{-1},\quad |C_{f_2}^\gamma|=0.21\,{\rm GeV}^{-1}.
\eeq
However, the relative sign of the couplings cannot be inferred and must be fitted to experiment. 

\section{Numerical results}
\label{sec:results}

\subsection{Sum rules}
\label{sec:res_sum_rule}

\begin{table}
\centering
\begin{tabular}{cccc}
\hline\hline
& full & $a\to\infty$ & no resonances\\\hline
$I^{(1)}$, CCL & $-0.62$ & $-1.15$ & $0.61$ \\
$I^{(1)}$, GKPRY & $-0.63$ & $-1.17$ & $0.60$ \\
$I^{(2)}$, CCL & $3.45$ & $3.58$ & $2.08$\\
$I^{(2)}$, GKPRY & $3.40$ & $3.53$ & $2.03$\\
\hline\hline
\end{tabular}
\caption{Integrals in the $I=2$ sum rule for the full left-hand cut, in the limit $a\to\infty$, and with resonances switched off.}
\label{table:sumrule_integrals} 
\end{table}

\begin{table}
\centering
\begin{tabular}{cccc}
\hline\hline
 & $\alpha_1-\beta_1$ & $\alpha_2-\beta_2$ & total\\\hline
ChPT  & $1.03\pm 0.14$ & $-4.29\pm 0.78$ & $0.18\pm 0.85$\\
GMM  & $0.80\pm 0.14$ & $-3.49\pm 0.60$ & $0.76\pm 0.68$\\
\hline\hline
\end{tabular}
\caption{Individual contribution to \eqref{sumrule1} from the polarizabilities (first two columns) and total value of the right-hand side of the sum rule (third column).}
\label{table:sumrule_ChPT_GMM} 
\end{table}

Before examining the reactions of interest, we turn to the numerical evaluation of the sum rules for $I=2$ derived in Sect.~\ref{sec:I2_sumrule}. As the $I=2$ $D$-wave $\pi\pi$ phase is very small, in practice no meaningful constraint results in these partial waves and we therefore restrict ourselves to the $S$-wave. These sum rules were written explicitly in \eqref{sumrule1} (see also App.~\ref{App:omnes_sol_I2}). 
The results  for $I^{(1)}$ and $I^{(2)}$ when the two different input $\pi \pi$ phases are chosen are shown in the first column of Table~\ref{table:sumrule_integrals}. The difference between using CCL and GKPRY $\pi\pi$ phases is very small in both cases. 

Evaluating the sum-rule integrals involves several approximations,
in particular, we have put $\Im h_{0,+}(t)$ to zero above the matching point, neglected partial waves with $J>2$, and used a resonance approximation for $\Im f_{J,\pm}(s)$. Therefore, we must now make sure that the dependence on the high-energy part of the integrals $I^{(1)}$ and $I^{(2)}$, higher partial waves, and details of the resonance description of the left-hand cut, is sufficiently small for the sum-rule constraint to be meaningful. We can estimate the accuracy of these approximations by sending the hyperbola parameter $a\to\infty$, because in the original setup the dispersion relation is independent of $a$. Thus, any residual dependence on $a$ provides a measure of the impact of the approximations made. The results of doing this are shown in the second column of Table~\ref{table:sumrule_integrals}: the once-subtracted integral depends strongly on the value of $a$, but the twice-subtracted version is already rather stable under $a\to\infty$. Doubling the effect of taking $a\to\infty$ to get a conservative estimate of the uncertainty in $I^{(2)}$, we conclude that
\beq
\label{I2}
I^{(2)}=3.45\pm 0.30.
\eeq

In order to further test the sensitivity of the sum rule to the modeling of the left-hand cut by a set of resonances we can check what happens if we switch off resonance contributions completely. 
In the case of $I^{(2)}$ even making this crude approximation entails a relatively modest shift in the result. As shown in the third column of Table~\ref{table:sumrule_integrals}, the resonances contribute less than $50\,\%$ to the full result, such that their contribution would have to be drastically wrong to exceed the error estimate given in~\eqref{I2}.

The stability of $I^{(2)}$ under these changes in high-energy input makes it worth taking \eqref{I2} seriously as a constraint on a particular linear combination of $(\alpha_1 - \beta_1)^{I=2}$ and $(\alpha_2 - \beta_2)^{I=2}$. Thus, we will now consider the resulting sum rule, which arises from the twice-subtracted Muskhelishvili--Omn\`es representation, in more detail.
 
First of all, we test if the parameter sets of Table~\ref{table:pol} fulfill the sum rule. The error analysis is complicated by the fact that in the GMM set no uncertainty estimate is given for ${(\alpha_1-\beta_1)^{\pi^{\pm}}}$, while in the ChPT set the error induced by the LEC dependence of $(\alpha_2-\beta_2)^{\pi^{\pm}}$ is difficult to assess. To obtain a rough estimate, we use the ChPT error for ${(\alpha_1-\beta_1)^{\pi^{\pm}}}$ also for GMM, and vice versa for $(\alpha_2-\beta_2)^{\pi^{\pm}}$. This, together with the number \eqref{I2}, leads to the results summarized in Table~\ref{table:sumrule_ChPT_GMM}. We conclude that the sum rule is fulfilled for both sets, although rather marginally in the case of GMM. (This is mainly due to the fact that ${(\alpha_2-\beta_2)^{\pi^0}}$ differs quite substantially between ChPT and GMM.) The largest uncertainty in the sum rule is driven by lack of knowledge of the quadrupole polarizability. 

Observing that both dipole polarizabilities as well as $(\alpha_2-\beta_2)^{\pi^0}$ have an accurate ChPT prediction, we can turn around the argument and use the sum rule to derive an improved value for $(\alpha_2-\beta_2)^{\pi^\pm}$. Using \eqref{I2} and the ChPT prediction for the isospin-two dipole polarizability, \eqref{sumrule1} leads to 
\beq
\label{polI2}
(\alpha_2-\beta_2)^{I=2}=(-18.2\pm 1.3)\cdot 10^{-4}{\rm fm}^5.
\eeq
Using, in addition, the ChPT prediction for ${(\alpha_2-\beta_2)^{\pi^0}}$, we find
\begin{align}
\label{ChPT_quad_pol}
(\alpha_2-\beta_2)^{\pi^\pm}&=(15.3\pm 3.7)\cdot 10^{-4}{\rm fm}^5,\notag\\
(\alpha_2-\beta_2)^{I=0}&=(39.4\pm 6.0)\cdot 10^{-4}{\rm fm}^5,
\end{align}
where the increase in uncertainty compared to \eqref{polI2} is due to the ChPT uncertainty in $(\alpha_2-\beta_2)^{\pi^0}$. In the remainder of the paper we will make use of the improved value \eqref{ChPT_quad_pol} when referring to the ChPT predictions for pion polarizabilities. Note that, as expected given the results of Table~\ref{table:sumrule_ChPT_GMM}, our sum-rule value of $(\alpha_2-\beta_2)^{\pi^\pm}$ is consistent with the first ChPT number quoted in Table~\ref{table:pol}, but it is not consistent with the larger number found when the LECs of \cite{Bijnens96} are taken as input.

\subsection{Total cross section}
\label{sec:total_cross_section}

\begin{figure}
\centering
\includegraphics[width=\linewidth,clip]{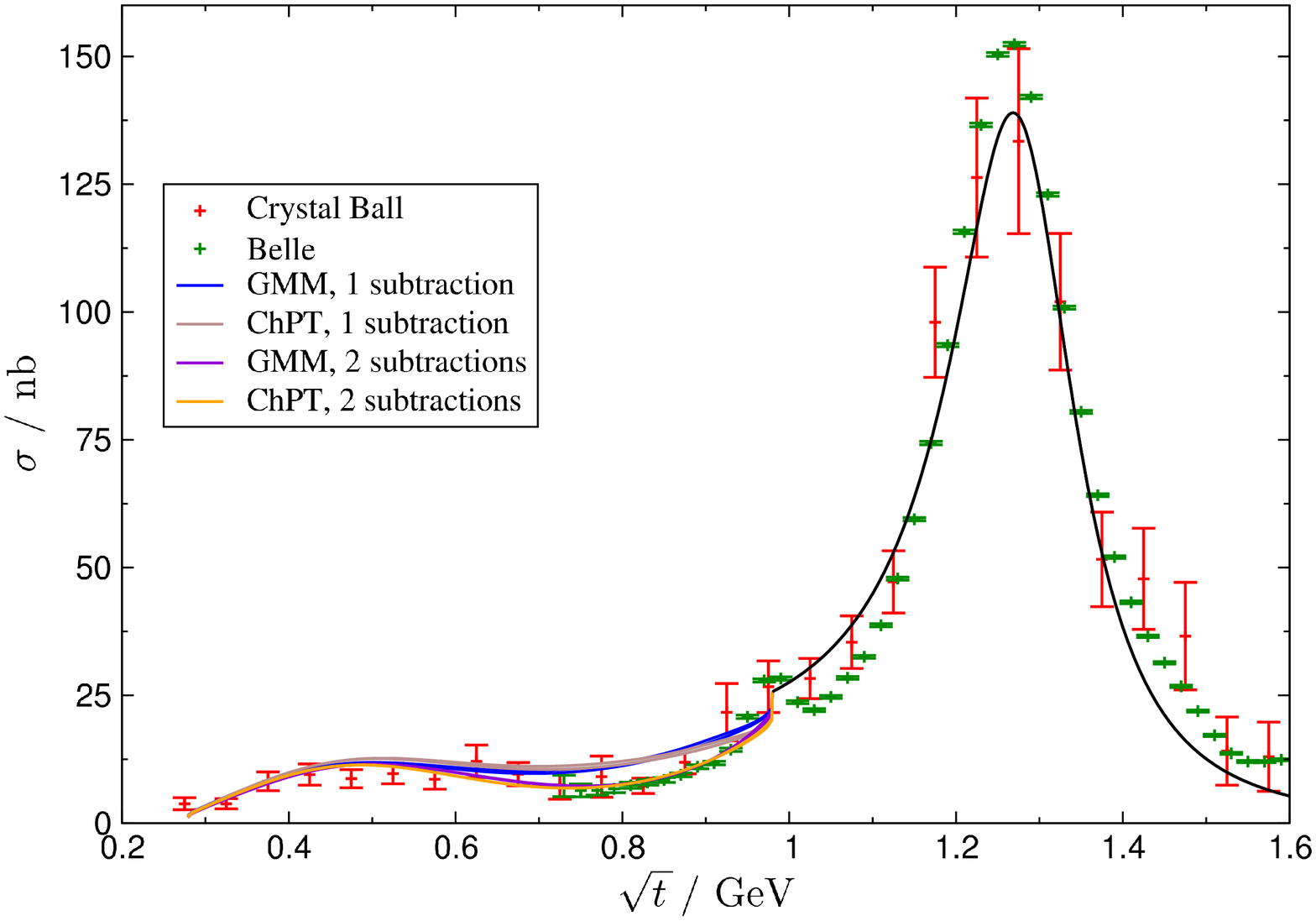}
\includegraphics[width=\linewidth,clip]{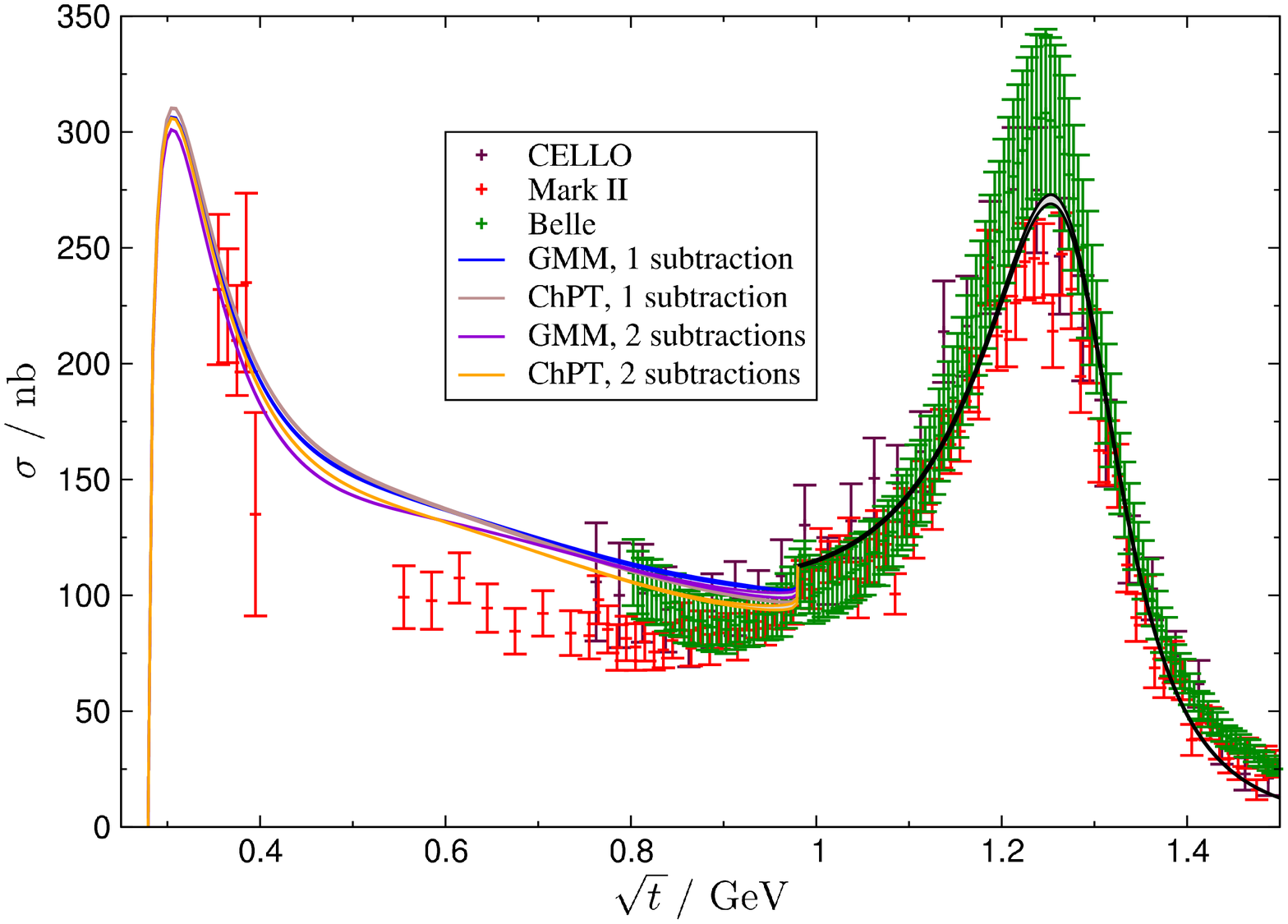}
\caption{Total cross section for $\gamma\gamma\to\pi^0\pi^0$ \cite{CB90,Belle09} and $\gamma\gamma\to\pi^+\pi^-$ \cite{Mark2,CELLO,Belle07} for $|\cos\theta|\leq 0.8$ and $|\cos\theta|\leq 0.6$, respectively.}
\label{fig:total_cross_section}
\end{figure}
\begin{figure}
\centering
\includegraphics[width=\linewidth,clip]{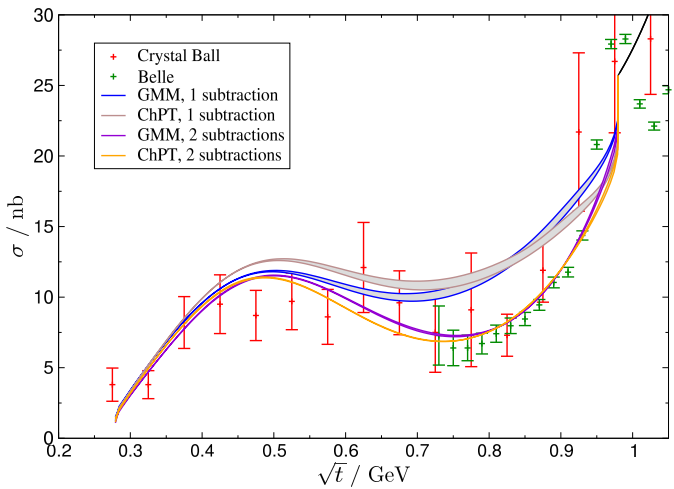}
\caption{Total cross section for $\gamma\gamma\to\pi^0\pi^0$ for $|\cos\theta|\leq 0.8$ in the low-energy region.}
\label{fig:total_cross_section_low_energy}
\end{figure}

Before performing the analytic continuation to the $\sigma$ pole, we wish to make sure that the amplitude on the real axis is reasonably well described---at least up to $\sqrt{t}=1\,{\rm GeV}$, which we assess to be the region which will influence the analytic continuation to the $\sigma$ pole. The results for the cross section are depicted in Figs.~\ref{fig:total_cross_section} and \ref{fig:total_cross_section_low_energy}. Below the matching point, the results for the once- and twice-subtracted formulation are provided for both ChPT and GMM polarizabilities. The uncertainty due to $\pi\pi$ input, represented by the grey band, is estimated by the variation between CCL and GKPRY phases and proves to be very small. The low-energy region  is totally dominated by the Born terms in the charged process, but it is very sensitive to the $\sigma$ in the neutral reaction. The prediction of the twice-subtracted dispersion relation is in especially good agreement with $\gamma\gamma\to\pi^0\pi^0$ data (see Fig.~\ref{fig:total_cross_section_low_energy}), with the level of agreement comparable to that obtained in the coupled-channel fit of~\cite{GM_10}.

Above the matching point, we exploit the fact that the cross section is dominated by the $f_2(1270)$, and thus can be well approximated by employing a Breit--Wigner description of this resonance in $h_{2,-}^{I=0}(t)$ and putting all other partial waves to zero. In this way,  \eqref{f2model} alone yields a good description of the neutral cross section above the matching point. In contrast, in the charged case an additional background is necessary. As observed in \cite{Drechsel:1999rf}, this can be most easily achieved by adding the Born terms and the off-shell contributions dropped in the transition from \eqref{f2_offshell} to \eqref{f2model} back into the charged-channel amplitude for $h_{2,-}(t)$. Moreover, after the transition to the isospin basis, we add a constant background phase to ensure matching with the $\pi\pi$ phase below the matching point. However, if $C_{f_2}^\pi C_{f_2}^\gamma$ is chosen to be negative, the mismatch of the phases is very small: we find a correction of $\delta_{\rm corr}=-0.09$ and $\delta_{\rm corr}=-0.04$ in order to obtain agreement with the CCL and GKPRY phases, respectively.

Finally, we comment on the analyticity properties of the partial waves at the matching point. As shown in the appendix of \cite{piK}, the solutions in terms of Omn\`es functions automatically fulfill continuity at the matching point, but the derivative at $\tm$ is not determined. Therefore, in general, strong cusps can occur at the matching point. For example, if the background in the charged reaction is dropped, the neutral cross section above $\tm$ is still correctly reproduced, but the input for the $I=0$ component changes, which affects the neutral cross section below $\tm$: the result for $|h_{2,-}^{I=0}(t)|$ exhibits a strong cusp below $\tm$, which translates into an (unphysical) sharp peak of about $15\,{\rm nb}$ in the neutral cross section directly below $\tm$. The fact that this effect is much smaller in the full solution provides evidence that our model for the high-energy region is reasonably accurate, because only a specific input of $\pi\pi$ phases, polarizabilities, and imaginary parts above $\tm$ will yield a smooth solution for $h_{2-}^{I=0}(t)$ around $t=\tm$. In the language of \cite{GasserWanders} such a solution corresponds to an ``analytic input''. If the input above the matching point were sufficiently well known, one could thus derive constraints on the polarizabilities by requiring a no-cusp condition. These constraints would be similar to those derived in \cite{ACGL,piK} for $\pi\pi$ and $\pi K$ scattering lengths. However, the input above the matching point is not very well known in $\gamma \gamma \rightarrow \pi \pi$, so we content ourself with requiring that the cusp at $\tm$ is not too large, such that the input we are using is reasonably close to being ``analytic''.

\subsection{Two-photon coupling of the $\boldsymbol{\sigma}$}

\begin{figure}
\centering
\includegraphics[width=\linewidth,clip]{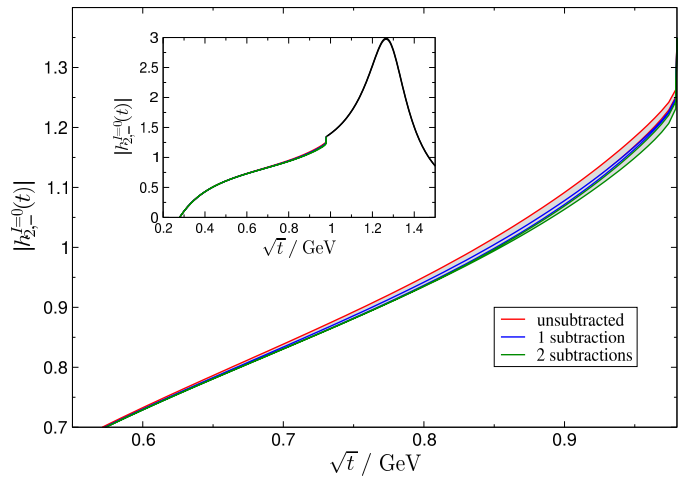}
\caption{Dependence of $|h_{2,-}^{I=0}(t)|$ on the number of subtractions. The grey bands indicate the difference between CCL and GKPRY phases.}
\label{fig:h2m_subtractions}
\end{figure}
\begin{figure}
\centering
\includegraphics[width=\linewidth,clip]{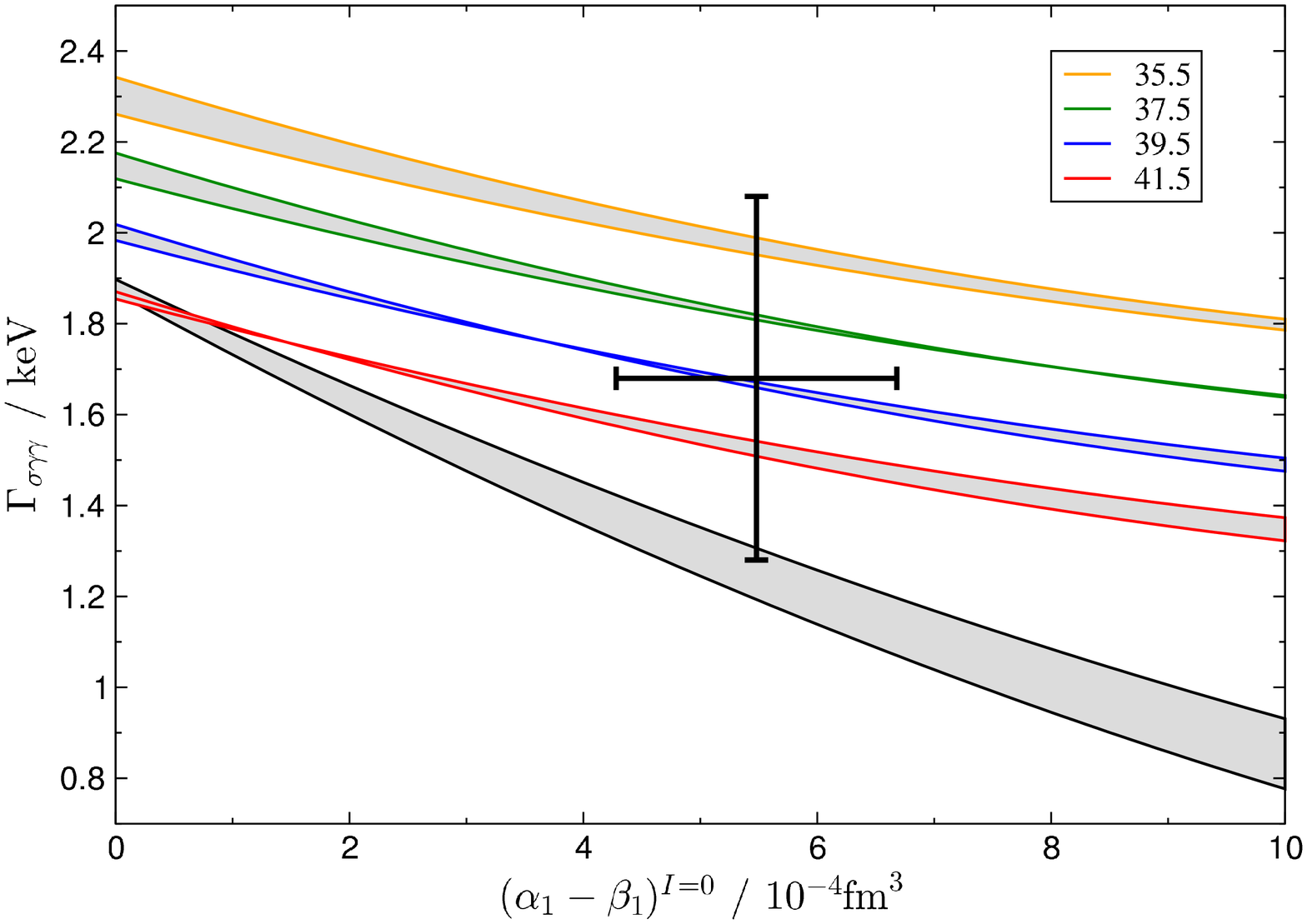}
\caption{$\Gamma_{\sigma\gamma\gamma}$ as a function of the $I=0$ pion polarizabilities. The black line refers to the unsubtracted case and the colored lines to the twice-subtracted version with $(\alpha_2-\beta_2)^{I=0}$ as indicated (in units of $10^{-4}{\rm fm}^5$). The grey band for the uncertainty in the $\pi\pi$ input is estimated by the variation found when CCL and GKPRY phases and $\sigma$ parameters are chosen. The cross corresponds to the twice-subtracted case plus ChPT input.}
\label{fig:width_pol}
\end{figure}
\begin{figure}
\centering
\includegraphics[width=\linewidth,clip]{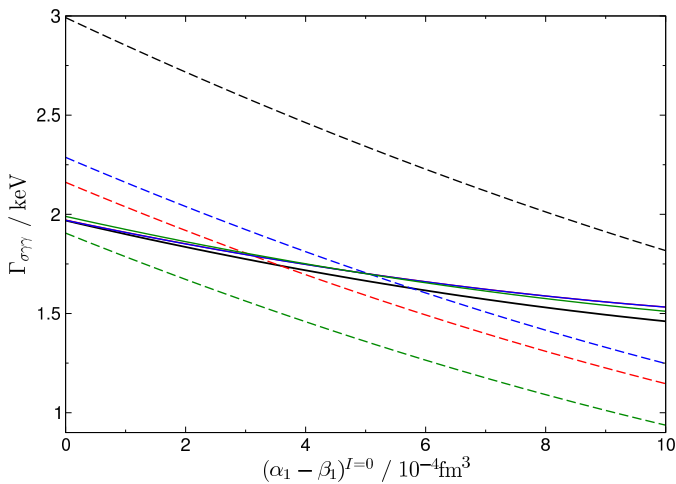}
\caption{Individual contributions to $\Gamma_{\sigma\gamma\gamma}$ for CCL phases and $\sigma$ parameters. The dashed lines refer to the once-subtracted version. The solid lines refer to the twice-subtracted version, with an input value of $(\alpha_2-\beta_2)^{I=0}$ from \eqref{ChPT_quad_pol}. The black lines denote the results for a left-hand cut modeled solely by the Born terms, while the red, blue, and green lines are obtained by successively adding resonances in the limit $a\to\infty$, terms for finite $a$, and $D$-wave contributions.}
\label{fig:width_pol_ind}
\end{figure}

We now present our results for the two-photon width $\Gamma_{\sigma\gamma\gamma}$ as a function of the pion polarizabilities. $(\alpha_1+\beta_1)$ and $(\alpha_2+\beta_2)$ only feature as subtraction constants in the $D$-waves, which, in turn, influence $\Gamma_{\sigma\gamma\gamma}$ only indirectly via the corresponding coupling to the $S$-wave in \eqref{SDwave}. Moreover, the imaginary part of these $D$-waves is dominated to a large extent by the $f_2(1270)$, and the dependence of $|h_{2,-}^{I=0}(t)|$ on the number of subtractions is very weak. Fig.~\ref{fig:h2m_subtractions} shows the results when we choose ChPT polarizabilities. The variation between the $|h_{2-}^{I=0}(t)|$ solutions for different numbers of subtractions is so small that the uncertainty in the $\pi\pi$ phases (estimated as the difference between CCL and GKPRY)  becomes of comparable size. Consequently, the dependence of $\Gamma_{\sigma\gamma\gamma}$ on $(\alpha_1 + \beta_1)^{I=0}$ and $(\alpha_2 + \beta_2)^{I=0}$ is totally negligible, such that we end up with the dipole polarizability $(\alpha_1-\beta_1)^{I=0}$ as the only free parameter that affects $\Gamma_{\sigma \gamma \gamma}$ in the once-subtracted dispersion relation. A second subtraction additionally requires the quadrupole polarizability $(\alpha_2-\beta_2)^{I=0}$ as input. The resulting correlation between $\Gamma_{\sigma\gamma\gamma}$ and the pion polarizabilities is depicted in Fig.~\ref{fig:width_pol}. This model-independent correlation is the main result of our study.

\begin{sloppypar}
The role of the different contributions to the left-hand cut is illustrated in Fig.~\ref{fig:width_pol_ind}: starting from the Born-term approximation (black line), we add resonances in the limit $a\to\infty$ (red line), the additional terms for finite $a=-7.5\,\mpi^2$ (blue line), and $D$-wave contributions (green line). The twice-subtracted version (solid lines) is hardly affected by any of these changes.  In the once-subtracted case (dashed lines) we see that $D$-wave and resonance contributions are of comparable size. We therefore expect that, as soon as resonances yield a significant contribution to the left-hand cut, the coupling between $S$- and $D$-waves should also become numerically important in  any description of data that is based on a Muskhelishvili--Omn\`es representation.
\end{sloppypar}
\begin{table}
\centering
\begin{tabular}{ccc}
\hline\hline
 & $1$ subtraction & $2$ subtractions\\\hline
ChPT & $1.3\pm 0.1$ & $1.7\pm 0.4$ \\
GMM & $1.4\pm 0.1$ & $2.0\pm 0.2$ \\
\hline\hline
\end{tabular}
\caption{Prediction for $\Gamma_{\sigma\gamma\gamma}$ in ${\rm keV}$ based on ChPT and GMM polarizabilities for CCL phases and $\sigma$ parameters.}
\label{table:width_pol} 
\end{table}

The results for $\Gamma_{\sigma\gamma\gamma}$ given ChPT and GMM choices for the polarizabilities are summarized in Table~\ref{table:width_pol}, where the errors only include the uncertainties from the pion polarizabilities. We have checked that these numbers are insensitive to the details of the input above the matching point.  For the ChPT parameters the results from the once- and twice-subtracted equations are consistent. However, there is significant tension between these two results in the case of GMM. This issue seems to be related to the relatively small value of $(\alpha_2-\beta_2)^{\pi^0}$ in that polarizability set: increasing this polarizability and thus bringing it closer to the ChPT prediction would both improve the fulfillment of the $I=2$ sum rule (cf.~Sect.~\ref{sec:res_sum_rule}) and bring $\Gamma_{\sigma\gamma\gamma}$ down from the $(2.0 \pm 0.2)$ keV result given in Table~\ref{table:width_pol}.

\begin{figure}
\centering
\includegraphics[width=\linewidth,clip]{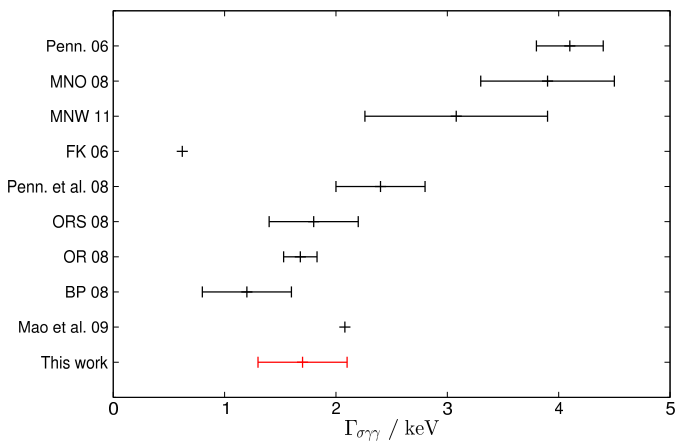}
\caption{Previous results for $\Gamma_{\sigma\gamma\gamma}$ with acronyms defined in Table~\ref{table:width_lit}. The red point corresponds to our result.}
\label{fig:width_lit}
\end{figure}

In view of Fig.~\ref{fig:width_pol_ind}, we consider the outcome of the twice-subtracted system of equations as the most reliable one, despite the fact that input for the quadrupole polarizability is required. Although the result based on the GMM parameters is ostensibly more precise, the Muskhelishvili--Omn\`es representation used in \cite{GM_10} is, for the reasons explained in Sect.~\ref{sec:comparison}, not fully consistent with the Roy--Steiner equations derived here. For this reason, we follow the philosophy of \cite{CGL} and combine the strict predictions of our Roy--Steiner equations with ChPT input for pion polarizabilities to obtain our final result
\beq
\Gamma_{\sigma\gamma\gamma}=(1.7\pm 0.4)\,{\rm keV},
\eeq   
which is depicted by the cross in Fig.~\ref{fig:width_pol}.
We note that the uncertainty here is broad enough to encompass all the central values in Table~\ref{table:width_pol}. A comparison with previous results for $\Gamma_{\sigma\gamma\gamma}$ is shown in Table~\ref{table:width_lit} and Fig.~\ref{fig:width_lit}.

\begin{table}
\centering
\begin{tabular}{clc}
\hline\hline
Reference& & $\Gamma_{\sigma\gamma\gamma}$\\\hline
Penn.~06 & \cite{Pennington06} & $4.1\pm 0.3$\\
MNO 08 & \cite{Mennessier:2008} & $3.9\pm 0.6$\\
MNW 11 & \cite{Mennessier:2010} & $3.08\pm 0.82$\\
FK 06 & \cite{FK_06} & $0.62$\\
Penn.~{\it et al.}~08 & \cite{Pennington08}, sol B & $2.4\pm 0.4$\\
ORS 08 & \cite{ORS_08} & $1.8\pm 0.4$\\
OR 08 & \cite{OR_08} & $1.68\pm 0.15$\\
BP 08 & \cite{Bernabeu_08} & $1.20\pm 0.40$\\
Mao {\it et al.}~09 & \cite{Mao09} & $2.08$\\
\hline\hline
\end{tabular}
\caption{Previous results for $\Gamma_{\sigma\gamma\gamma}$ in ${\rm keV}$.}
\label{table:width_lit} 
\end{table}

\section{Summary and conclusion}
\label{sec:conlusion}

In this paper we constructed a full set of Roy--Steiner equations for $\gamma\pi\to\gamma\pi$. In particular, we worked out all necessary integral kernels for zero, one, and two subtractions explicitly up to $D$-waves, and directly identified the subtraction constants with the pion polarizabilities. We studied the range of validity of the system, and found that the equations for the crossed channel $\gamma\gamma\to\pi\pi$ are rigorously valid up to $1\,{\rm GeV}$---a domain that comfortably includes the $\sigma$ pole. Extending this range requires additional assumptions. Truncating the system at $J=2$, we then concentrated on the equations for $\gamma\gamma\to\pi\pi$, whose solution in terms of a Muskhelishvili--Omn\`es representation with a finite matching point was discussed. Comparing our equations with existing approaches in the literature, we found a coupling between $S$- and $D$-waves, which has previously been neglected in calculations of $\gamma \gamma \to \pi \pi$ based on dispersion relations, but seems to be numerically comparable to the contributions of resonances in $\gamma \gamma \to \pi \pi$ to the left-hand cut. 

\begin{sloppypar}
Demanding continuity of the Muskhelishvili--Omn\`es representation at the matching point, we derived sum rules for the $I=2$ partial waves that relate dipole and quadrupole polarizabilities to integrals over the left-hand cut. We used the $S$-wave sum rule, together with ChPT input for the neutral-pion quadrupole polarizability and the dipole polarizabilities, to obtain a new, more accurate, prediction for the charged-pion quadrupole polarizability
\beq
(\alpha_2-\beta_2)^{\pi^\pm}=(15.3\pm 3.7)\cdot 10^{-4}{\rm fm}^5.
\eeq
While the central value is hardly shifted, the error estimate is difficult to obtain in ChPT alone due to a strong dependence on poorly known low-energy constants.
\end{sloppypar}

The main application of our formalism though, was a study of the two-photon width of the $\sigma$ resonance. To this end, we first showed that the cross section for both the charged and neutral channel can be accurately reproduced by approximating the high-energy region above $1\,{\rm GeV}$ by a Breit--Wigner ansatz for the $f_2(1270)$ resonance and employing a suitably chosen background amplitude in the charged case. With this input, our Muskhelishvili--Omn\`es representation yields a good description of the available data in the low-energy region.  

We then presented the results of the analytic continuation to the $\sigma$ pole as a correlation plot between the pertinent pion polarizabilities and $\Gamma_{\sigma\gamma\gamma}$. Our most general finding is a correlation 
between $I=0$ pion polarizabilities and $\Gamma_{\sigma\gamma\gamma}$. This correlation results solely from analyticity, unitarity, crossing symmetry, and the accurately known $\pi \pi$ phases in the region below $1\,{\rm GeV}$. We also provided a specific result for $\Gamma_{\sigma\gamma\gamma}$ by taking ChPT predictions for the pertinent pion polarizabilities. We therefore look forward to the results of the ongoing efforts at COMPASS to extract pion polarizabilities, which---as shown here---will further improve knowledge of $\Gamma_{\sigma\gamma\gamma}$.

\begin{acknowledgement}

\begin{sloppypar}
\textbf{Acknowledgements\ }
We thank Irinel Caprini, Gilberto Colangelo, Bachir Moussallam, and Jos\'e Pel\'aez for useful correspondence, in particular Gilberto Colangelo and Jos\'e Pel\'aez for communicating the results of \cite{Bern} and \cite{EGKP10} prior to publication, and Bastian Kubis and Bachir Moussallam 
 for carefully reading the manuscript. 
This research was supported by the DFG (SFB/TR 16, ``Subnuclear Structure of Matter''), the program ``Kurzstipendien f\"ur DoktorandInnen'' of the DAAD, the Bonn-Cologne Graduate School of Physics and Astronomy, the US Department of Energy (Office of Nuclear Physics, under contract No.~DE-FG02-93ER40756 with Ohio University), and CONICET. M.~H. and C.~S. would like to thank the Institute of Nuclear and Particle Physics at Ohio University for its hospitality during a visit where most of this work was done.
\end{sloppypar}
\end{acknowledgement}

\begin{appendix}

\renewcommand{\theequation}{\thesection.\arabic{equation}}

\begin{strip}

\setcounter{equation}{0}
\section{Kernel functions for the $\boldsymbol{s}$-channel projection}
\label{app:kernel_schannel}

The functions $N_J^\pm(s)$ for $J\leq 2$ are given by
\begin{align}
 N_1^+(s)&=\frac{s}{\mpi^2-s}-\frac{\mpi^2}{\qq^2}\Big\{d_{11}^1(y)Q_0(y)-\frac{1}{2}\Big\}+\Delta N_1^+(s),\notag\\
N_2^+(s)&=-\frac{1}{3}\frac{s}{\mpi^2-s}-\frac{\mpi^2}{\qq^2}\Big\{d_{11}^2(y)Q_0(y)-\frac{1}{2}-y\Big\}+\Delta N_2^+(s),\notag\\
 N_1^-(s)&=-\frac{\mpi^2}{\mpi^2-s}+\frac{\mpi^2}{\qq^2}\Big\{d_{1,-1}^1(y)Q_0(y)+\frac{1}{2}\Big\}+\Delta N_1^-(s),\notag\\
N_2^-(s)&=-\frac{1}{3}\frac{\mpi^2}{\mpi^2-s}+\frac{\mpi^2}{\qq^2}\Big\{d_{1,-1}^2(y)Q_0(y)-\frac{1}{2}+y\Big\}+\Delta N_2^-(s),
\end{align}
where
\beq
y=-\frac{s+\mpi^2}{s-\mpi^2},
\eeq
and
\beq
Q_0(z)=\frac{1}{2}\int\limits_{-1}^1\frac{\diff x}{z-x},\quad
Q_0(z\pm i\eps)=\frac{1}{2}\log\bigg|\frac{1+z}{1-z}\bigg|\mp i\frac{\pi}{2}\theta(1-z^2),
\eeq
denotes the lowest Legendre polynomial of the second kind. The remainders $\Delta N_J^\pm(s)$ 
\begin{align}
\Delta N_1^{+\,(0)}(s)&=0,\quad  \Delta N_1^{+\,(1)}(s)=\frac{2 s}{3\mpi\alpha}(\alpha_1+\beta_1)\qq^2, \quad
\Delta N_1^{+\,(2)}(s)=\Delta N_1^{+\,(1)}(s)-\frac{s}{18\mpi \alpha }(\alpha_2+\beta_2)\qq^4,\notag\\
\Delta N_2^{+\,(0)}(s)&=\Delta N_2^{+\,(1)}(s)=0,\quad
\Delta N_2^{+\,(2)}(s)=\frac{s}{30\mpi \alpha }(\alpha_2+\beta_2)\qq^4,\notag\\
\Delta N_1^{-\,(0)}(s)&=-\frac{2\qq^2}{3(\mpi^2-a)},\quad
\Delta N_1^{-\,(1)}(s)=\frac{2\mpi}{3\alpha}(\alpha_1-\beta_1)\qq^2
+\frac{1}{2\mpi\alpha }(\alpha_1+\beta_1)\qq^4,\notag\\
\Delta N_1^{-\,(2)}(s)&=\frac{2\mpi}{3\alpha}(\alpha_1-\beta_1)\qq^2-\frac{\mpi}{6\alpha}(\alpha_2-\beta_2)\qq^4
-\frac{2}{15\mpi\alpha}(\alpha_2+\beta_2)\qq^6,\\
\Delta N_2^{-\,(0)}(s)&=0,\quad \Delta N_2^{-\,(1)}(s)=-\frac{1}{10\mpi\alpha}(\alpha_1+\beta_1)\qq^4,\quad
\Delta N_2^{-\,(2)}(s)=\frac{\mpi}{30\alpha}(\alpha_2-\beta_2)\qq^4+\frac{2}{45\mpi\alpha}(\alpha_2+\beta_2)\qq^6,\notag
\end{align}
contain the pion polarizabilities according to the number of subtractions indicated by the superscript.

\subsection{$\boldsymbol{s}$-channel}

The kernels for the unsubtracted case read

\begin{align}
 K_{11}^{++}(s,s')&=\frac{s\, \qq^2}{s' \qq'^2}\Bigg\{\frac{1}{s'-s}-\frac{1}{s'-a}-\frac{3}{2\qq^2}\bigg[\frac{(1+x_s)^2}{4}Q_0(x_s)-\frac{2+x_s}{4}\bigg]\Bigg\},\notag\\
K_{12}^{++}(s,s')&=\frac{s\, \qq^2}{s' \qq'^2}\Bigg\{\bigg(\frac{1}{s'-s}-\frac{1}{s'-a}\bigg)\frac{5}{3}(2\beta +\alpha -1)\notag\\
&-\frac{5}{2\qq^2}\bigg[\frac{(1+x_s)^2}{4}(2x_s'-1)Q_0(x_s)-\Big(\frac{2}{3}\alpha +\beta -\frac{1}{2}\Big)-\Big(\alpha +\frac{\beta }{2}-\frac{1}{4}\Big) x_s-\frac{\alpha  }{2}x_s^2\bigg]\Bigg\},\notag\\
K_{21}^{++}(s,s')&=-\frac{s}{s' \qq'^2}\frac{3}{2}\bigg\{\frac{(1+x_s)^2}{4}(2x_s-1)Q_0(x_s)-\frac{1}{6}-\frac{3 x_s}{4}-\frac{x_s^2}{2}\bigg\},\notag\\
K_{22}^{++}(s,s')&=\frac{s\,\qq^2}{s' \qq'^2}\Bigg\{\bigg(\frac{1}{s'-s}-\frac{1}{s'-a}\bigg)\alpha-\frac{5}{2\qq^2}\bigg[\frac{(1+x_s)^2}{4}(2x_s-1)(2x_s'-1)Q_0(x_s)\notag\\
&-\Big(\frac{\beta }{3}-\frac{1}{6}\Big)-\Big(\frac{\alpha }{3}+\frac{3 }{2}\beta -\frac{3}{4}\Big) x_s-\Big(\beta +\frac{3 }{2}\alpha -\frac{1}{2}\Big) x_s^2-\alpha  x_s^3\bigg]\Bigg\},\notag\\
K_{11}^{-+}(s,s')&=\frac{\qq^2}{2s' \qq'^2}\Bigg\{\bigg(\frac{1}{s'-s}-\frac{1}{s'-a}\bigg)\Big(\frac{3}{2}\qq^2-\frac{1}{2}\qq'^2(\alpha -2\beta +2)\Big)-\frac{s'-s}{2(s'-a)}\Bigg\},\notag\\
K_{12}^{-+}(s,s')&=\frac{\qq^2}{2s' \qq'^2}\Bigg\{\bigg(\frac{1}{s'-s}-\frac{1}{s'-a}\bigg)\Big\{\Big(5 \beta -3 \alpha -\frac{5}{2}\Big)\qq^2+\frac{\qq'^2}{6} \left(8 \alpha ^2+5 \alpha  (3-4 \beta)+10 \left(1-3 \beta +2 \beta ^2\right)\right)\Big\}\notag\\
&+\frac{5}{6}\frac{s'-s}{s'-a}(1+\alpha-2\beta)\Bigg\},\notag\\
K_{21}^{-+}(s,s')&=-\frac{\qq^2}{2s' \qq'^2}\bigg(\frac{1}{s'-s}-\frac{1}{s'-a}\bigg)\frac{3}{10}(\qq^2-\qq'^2 \alpha ),\notag\\
K_{22}^{-+}(s,s')&=\frac{\qq^2}{2s' \qq'^2}\Bigg\{\bigg(\frac{1}{s'-s}-\frac{1}{s'-a}\bigg)\Big(\Big(\frac{1}{2}+\frac{5 \alpha }{3}-\beta \Big)\qq^2-\qq'^2 \frac{\alpha }{6} (9+4 \alpha -12 \beta )\Big)-\frac{\alpha}{2}\frac{s'-s}{s'-a}\Bigg\},\notag\\
K_{11}^{--}(s,s')&=\frac{\qq^2}{\qq'^2}\Bigg\{\frac{1}{s'-s}-\frac{1}{s'-a}-\frac{3}{2\qq^2}\bigg[\frac{(1-x_s)^2}{4}Q_0(x_s)+\frac{2-x_s}{4}\bigg]\Bigg\},\notag\\
K_{12}^{--}(s,s')&=\frac{\qq^2}{\qq'^2}\Bigg\{\bigg(\frac{1}{s'-s}-\frac{1}{s'-a}\bigg)\frac{5}{3} (2 \beta -\alpha +1)-\frac{5}{2\qq^2}\bigg[\frac{(1-x_s)^2}{4}(2x_s'+1)Q_0(x_s)\notag\\
&-\Big(\frac{2 }{3}\alpha -\beta -\frac{1}{2}\Big)-\Big(\frac{\beta }{2}-\alpha +\frac{1}{4}\Big) x_s-\frac{\alpha  }{2}x_s^2\bigg]\Bigg\},\notag\\
K_{21}^{--}(s,s')&=-\frac{1}{\qq'^2}\frac{3}{2}\Bigg\{\frac{(1-x_s)^2}{4}(2x_s+1)Q_0(x_s)-\frac{1}{6}+\frac{3 x_s}{4}-\frac{x_s^2}{2}\Bigg\},\notag\\
K_{22}^{--}(s,s')&=\frac{\qq^2}{\qq'^2}\Bigg\{\bigg(\frac{1}{s'-s}-\frac{1}{s'-a}\bigg)\alpha -\frac{5}{2\qq^2}\bigg[\frac{(1-x_s)^2}{4}(2x_s+1)(2x_s'+1)Q_0(x_s)\notag\\
&-\Big(\frac{\beta }{3}+\frac{1}{6}\Big)-\Big(\frac{\alpha }{3}-\frac{3 }{2}\beta -\frac{3}{4}\Big) x_s-\Big(\beta -\frac{3 }{2}\alpha +\frac{1}{2}\Big) x_s^2-\alpha  x_s^3\bigg]\Bigg\},
\end{align}
where
\beq
\label{zs_zsprime}
\alpha=\frac{\qq^2}{\qq'^2}\frac{s-a}{s'-a},\quad \beta=1-\alpha-\frac{s'-s}{s'-a}\frac{s+s'-2\mpi^2}{2\qq'^2},\quad z_s'=\alpha z_s+\beta,\quad
x_s=1-\frac{s+s'-2\mpi^2}{2\qq^2},\quad x_s'=\alpha x_s+\beta.
\eeq
The corresponding versions for the once-($^{(1)}$) and twice-($^{(2)}$) subtracted case can be obtained by adding
\begin{align}
 \Delta K_{JJ'}^{++\,(1)}(s,s')&=-\frac{s\,\qq^2}{s'\qq'^2}(2J'+1)\frac{2}{3}\delta_{J1}\bigg(\frac{2}{s'-\mpi^2}-\frac{1}{s'-a}\bigg)\frac{d_{11}^{J'}(z_s')}{1+z_s'}\bigg|_0,\notag\\
\Delta K_{JJ'}^{++\,(2)}(s,s')&=\frac{2s\,\qq^4}{s'\qq'^2}(2J'+1)\Big(\frac{\delta_{J1}}{3}-\frac{\delta_{J2}}{5}\Big)
\bigg\{\bigg(\frac{2}{s'-\mpi^2}-\frac{1}{s'-a}\bigg)\partial_t\frac{d_{11}^{J'}(z_s')}{1+z_s'}\bigg|_0-\frac{1}{(s'-\mpi^2)^2}\frac{d_{11}^{J'}(z_s')}{1+z_s'}\bigg|_0\bigg\},\notag\\
\Delta K_{JJ'}^{-+\,(1)}(s,s')&=-\frac{\qq^2}{4s'\qq'^2}(2J'+1)\bigg(\frac{2}{s'-\mpi^2}-\frac{1}{s'-a}\bigg)\frac{d_{11}^{J'}(z_s')}{1+z_s'}\bigg|_0
\Big\{-2\qq'^2\big(1-z_s'\big|_0\big)\frac{2}{3}\delta_{J1}+2\qq^2\Big(\delta_{J1}-\frac{\delta_{J2}}{5}\Big)\Big\},\notag\\
\Delta K_{JJ'}^{-+\,(2)}(s,s')&=-\frac{\qq^4}{2s'\qq'^2}(2J'+1)\Bigg\{\bigg[-2\qq^2\Big(\frac{8}{5}\delta_{J1}-\frac{8}{15}\delta_{J2}+\frac{8}{105}\delta_{J3}\Big)-4\mpi^2\Big(\delta_{J1}-\frac{\delta_{J2}}{5}\Big)\bigg]\notag\\
&\times\bigg\{\bigg(\frac{2}{s'-\mpi^2}-\frac{1}{s'-a}\bigg)\partial_t\frac{d_{11}^{J'}(z_s')}{1+z_s'}\bigg|_0-\frac{1}{(s'-\mpi^2)^2}\frac{d_{11}^{J'}(z_s')}{1+z_s'}\bigg|_0\bigg\}\notag\\
&\hspace{-55pt}-\Big(\delta_{J1}-\frac{\delta_{J2}}{5}\Big)\bigg\{\bigg(\frac{2}{s'-\mpi^2}-\frac{1}{s'-a}\bigg)\partial_t(t'-4\mpi^2)\frac{d_{11}^{J'}(z_s')}{1+z_s'}\bigg|_0-\frac{1}{(s'-\mpi^2)^2}(t'-4\mpi^2)\frac{d_{11}^{J'}(z_s')}{1+z_s'}\bigg|_0\bigg\}
\Bigg\},\notag\\
\Delta K_{JJ'}^{--\,(1)}(s,s')&=-\frac{\qq^2}{\qq'^2}(2J'+1)\frac{2}{3}\delta_{J1}\bigg(\frac{2}{s'-\mpi^2}-\frac{1}{s'-a}\bigg)\frac{d_{1,-1}^{J'}(z_s')}{1-z_s'}\bigg|_0,\notag\\
\Delta K_{JJ'}^{--\,(2)}(s,s')&=\frac{2\qq^4}{\qq'^2}(2J'+1)\Big(\delta_{J1}-\frac{\delta_{J2}}{5}\Big)
\bigg\{\bigg(\frac{2}{s'-\mpi^2}-\frac{1}{s'-a}\bigg)\partial_t\frac{d_{1,-1}^{J'}(z_s')}{1-z_s'}\bigg|_0-\frac{1}{(s'-\mpi^2)^2}\frac{d_{1,-1}^{J'}(z_s')}{1-z_s'}\bigg|_0\bigg\},
\end{align}
with
\beq
z_s'\big|_0=-\frac{s'+a}{s'-a},\quad \partial_t z_s'\big|_0=\frac{\mpi^2-a}{2\qq'^2(s'-a)}.
\eeq

\subsection{$\boldsymbol{t}$-channel}

The non-vanishing unsubtracted kernel functions are
\begin{align}
 G^{+-}_{12}(s,t')&=\frac{5\sqrt{6}\,s}{2t'\left(t'-4\mpi^2\right)}\Big\{(1+x_t)^2Q_0(x_t)-2-x_t\Big\},\notag\\
G^{+-}_{22}(s,t')&=\frac{5\sqrt{6}\,s}{2t'\left(t'-4\mpi^2\right)}\Big\{(1+x_t)^2(2x_t-1)Q_0(x_t)-\frac{2}{3}-3 x_t-2 x_t^2\Big\},\notag\\
G^{-+}_{10}(s,t')&=\frac{1}{2t'}\Big\{(1-x_t)^2Q_0(x_t)+2-x_t\Big\},\notag\\
G^{-+}_{20}(s,t')&=\frac{1}{2t'}\Big\{(1-x_t)^2(2x_t+1)Q_0(x_t)-\frac{2}{3}+3 x_t-2 x_t^2\Big\},\notag\\
G^{-+}_{12}(s,t')&=\frac{5}{2t'}\Big\{(1-x_t)^2P_2(x_t')Q_0(x_t)-(1+2 \gamma -3 \delta )+\Big(\frac{1}{2}+3 \gamma -\frac{3 \delta }{2}\Big) x_t-\frac{3 \gamma  x_t^2}{2}\Big\},\notag\\
G^{-+}_{22}(s,t')&=\frac{5}{2t'}\Big\{(1-x_t)^2(2x_t+1)P_2(x_t')Q_0(x_t)
-\Big(\delta -\frac{1}{3}\Big)-\Big(\gamma -\frac{9}{2} \delta +\frac{3}{2}\Big) x_t
-\Big(3 \delta -\frac{9 }{2}\gamma -1\Big) x_t^2-3 \gamma  x_t^3\Big\},\notag\\
G^{--}_{12}(s,t')&=\frac{5\sqrt{6}\,\qq^2}{3t'\left(t'-4\mpi^2\right)},\quad G^{--}_{22}(s,t')=0,
\end{align}
where
\begin{align}
 \gamma&=\frac{8\qq^2(s-a)}{t'(t'-4\mpi^2)},\quad \delta=\frac{(t'-2\mpi^2+2a)^2-4(s-a)(2\qq^2+2\mpi^2-s-a)}{t'(t'-4\mpi^2)},\quad z_t'=\sqrt{\gamma z_s+\delta},\notag\\
x_t&=1+\frac{t'}{2\qq^2},\quad x_t'=\sqrt{\gamma x_t+\delta},
\end{align}
and the subtracted versions are obtained by adding
\begin{align}
\Delta G_{JJ'}^{+-\,(1)}(s,t')&=-\frac{8s\,\qq^2}{t'^2(t'-4\mpi^2)}(2J'+1)\frac{2}{3}\delta_{J1}\frac{d_{20}^{J'}(z_t')}{1-z_t'^2}\bigg|_0,\notag\\
\Delta G_{JJ'}^{+-\,(2)}(s,t')&=\frac{16s\,\qq^4}{t'^2(t'-4\mpi^2)}(2J'+1)\Big(\frac{\delta_{J1}}{3}-\frac{\delta_{J2}}{5}\Big)\bigg\{\frac{1}{t'}\frac{d_{20}^{J'}(z_t')}{1-z_t'^2}\bigg|_0+\partial_t\frac{d_{20}^{J'}(z_t')}{1-z_t'^2}\bigg|_0\bigg\},\notag\\
\Delta G_{JJ'}^{-+\,(1)}(s,t')&=-\frac{2\qq^2}{t'^2}(2J'+1)\frac{2}{3}\delta_{J1}P_{J'}(z_t')\big|_0,\notag\\
\Delta G_{JJ'}^{-+\,(2)}(s,t')&=\frac{4\qq^4}{t'^2}(2J'+1)\Big(\delta_{J1}-\frac{\delta_{J2}}{5}\Big)\bigg\{\partial_tP_{J'}(z_t')\big|_0+\frac{P_{J'}(z_t')\big|_0}{t'}\bigg\},\notag\\
\Delta G_{JJ'}^{--\,(1)}(s,t')&=-\frac{2\qq^2}{t'^2(t'-4\mpi^2)}(2J'+1)\Big\{t'\frac{2}{3}\delta_{J1}+2\qq^2\Big(\delta_{J1}-\frac{\delta_{J2}}{5}\Big)\Big\}\frac{d_{20}^{J'}(z_t')}{1-z_t'^2}\bigg|_0,\notag\\
\Delta G_{JJ'}^{--\,(2)}(s,t')&=\frac{4\qq^4}{t'^2(t'-4\mpi^2)}(2J'+1)\bigg[t'\Big(\delta_{J1}-\frac{\delta_{J2}}{5}\Big)+2\qq^2\Big(\frac{8}{5}\delta_{J1}-\frac{8}{15}\delta_{J2}+\frac{8}{105}\delta_{J3}\Big)\bigg]\notag\\
&\times\bigg\{\frac{1}{t'}\frac{d_{20}^{J'}(z_t')}{1-z_t'^2}\bigg|_0+\partial_t\frac{d_{20}^{J'}(z_t')}{1-z_t'^2}\bigg|_0\bigg\},
\end{align}
with
\beq
z_t'^2\big|_0=1+\frac{4a}{t'-4\mpi^2},\quad \partial_t z_t'^2\big|_0=\frac{4(\mpi^2-a)}{t'(t'-4\mpi^2)}.
\eeq

\setcounter{equation}{0}
\section{Kernel functions for the $\boldsymbol{t}$-channel projection}
\label{app:kernel_tchannel}

The contributions from Born terms and subtraction constants for $J\leq 2$ are
\begin{align}
\label{tchannel_Born}
 \tilde N_0^+(t)&=\frac{8\mpi^2}{\sigma(t)t}Q_0\Big(\frac{1}{\sigma(t)}\Big)+\Delta\tilde N_0^{+}(t),\quad
\tilde N_2^+(t)=\frac{8\mpi^2}{\sigma(t)t}\bigg\{P_2\Big(\frac{1}{\sigma(t)}\Big)Q_0\Big(\frac{1}{\sigma(t)}\Big)-\frac{3}{2\sigma(t)}\bigg\},\notag\\
\tilde N_2^-(t)&=-\frac{8\mpi^2}{\sigma(t)t}\bigg\{d_{20}^2\Big(\frac{1}{\sigma(t)}\Big)Q_0\Big(\frac{1}{\sigma(t)}\Big)+\frac{\sqrt{6}}{4\sigma(t)}\bigg\}
+\frac{2}{\sqrt{6}}+\Delta\tilde N_2^{-}(t),\notag\\
\Delta\tilde N_0^{+\,(0)}(t)&=-\frac{t}{2(\mpi^2-a)},\quad
\Delta\tilde N_0^{+\,(1)}(t)=\frac{\mpi}{2\alpha}(\alpha_1- \beta_1)t-\frac{1}{8\mpi\alpha }(\alpha_1+ \beta_1)t^2,\notag\\
\Delta\tilde N_0^{+\,(2)}(t)&=\frac{\mpi}{2\alpha}(\alpha_1- \beta_1)t+\frac{\mpi}{24\alpha}(\alpha_2- \beta_2)t^2
-\frac{1}{96\mpi\alpha}(\alpha_2+\beta_2)t^3,\\
\Delta\tilde N_2^{-\,(0)}(t)&=0,\quad
\Delta\tilde N_2^{-\,(1)}(t)=\frac{t(t-4\mpi^2)}{5\sqrt{6}\,\mpi^2}\frac{\mpi}{2\alpha}(\alpha_1+ \beta_1),\quad
\Delta\tilde N_2^{-\,(2)}(t)=\frac{t(t-4\mpi^2)}{5\sqrt{6}\,\mpi^2}\frac{\mpi}{2\alpha}\Big(\alpha_1+ \beta_1+\frac{t}{12}(\alpha_2+\beta_2)\Big).\notag
\end{align}

\subsection{$\boldsymbol{s}$-channel}

The non-vanishing kernel functions for the unsubtracted case are
\begin{align}
\tilde G^{++}_{01}(t,s')&=\frac{3t}{16s' \qq'^2(s'-a)}\Big\{2\qq'^2\frac{3 \tilde \gamma+3 \tilde \delta-1}{3 \tilde \gamma}-2(s'-\mpi^2)\Big\},\notag\\
\tilde G^{++}_{02}(t,s')&=\frac{5t}{16s' \qq'^2(s'-a)}\bigg\{2\qq'^2\frac{20 \tilde \delta-6-15 \left(\tilde \gamma(\tilde \gamma-1)+3 \tilde \gamma \tilde \delta+2 \tilde \delta^2\right)}{15 \tilde \gamma^2}+\frac{2(s'-\mpi^2)}{\tilde\gamma}\Big(\tilde\gamma+2\tilde\delta-\frac{2}{3}\Big)\bigg\},\notag\\
\tilde G^{++}_{21}(t,s')&=-\frac{t}{20s'(s'-a)\tilde \gamma},\notag\\
\tilde G^{++}_{22}(t,s')&=\frac{5t}{16s' \qq'^2(s'-a)}\bigg\{2\qq'^2\frac{2 (-12+21 \tilde \gamma+28 \tilde \delta)}{105 \tilde \gamma^2}-\frac{s'-\mpi^2}{15\qq'^2(s'-a)}t(t-4\mpi^2)\bigg\},\notag\\
\tilde G^{+-}_{01}(t,s')&=-\frac{3t}{4 \qq'^2(s'-a)}+\frac{3t}{4 \qq'^2p_tq_t}Q_0(\tilde x_t),\notag\\
\tilde G^{+-}_{02}(t,s')&=-\frac{5t}{4 \qq'^2(s'-a)}\frac{2+3 \tilde \gamma-6 \tilde \delta}{3 \tilde \gamma}+\frac{5t}{4 \qq'^2p_tq_t}\bigg\{\left(2\tilde x_t'+1\right)Q_0(\tilde x_t)-\frac{2 \tilde x_t}{\tilde \gamma}\bigg\},\notag\\
\tilde G^{+-}_{21}(t,s')&=\frac{3t}{4 \qq'^2p_tq_t}\Big\{P_2(\tilde x_t)Q_0(\tilde x_t)-\frac{3}{2}\tilde x_t\Big\},\notag\\
\tilde G^{+-}_{22}(t,s')&=-\frac{t}{3\qq'^2(s'-a)\tilde \gamma}+\frac{5t}{4 \qq'^2p_tq_t}\bigg\{\left(2\tilde x_t'+1\right)P_2(\tilde x_t)Q_0(\tilde x_t)-\Big(\frac{3}{2}-\frac{3 \tilde \delta}{\tilde \gamma}\Big) \tilde x_t-\frac{3 \tilde x_t^3}{\tilde \gamma}\bigg\},\notag\\
\tilde G^{-+}_{21}(t,s')&=-\frac{t\left(t-4\mpi^2\right)}{s' \qq'^2(s'-a)}\frac{3}{20\sqrt{6}}+\frac{3p_tq_t}{s' \qq'^2}\bigg\{\left(1-\tilde x_t^2\right)d_{20}^2(\tilde x_t)Q_0(\tilde x_t)+\frac{\tilde x_t}{2 \sqrt{6}}\left(5-3\tilde x_t^2\right)\bigg\},\notag\\
\tilde G^{-+}_{22}(t,s')&=\frac{t(t-4\mpi^2)}{s' \qq'^2(s'-a)}\frac{7 \tilde \gamma+14 \tilde \delta-2}{28\sqrt{6}\,\tilde \gamma}+\frac{5p_tq_t}{s' \qq'^2}\bigg\{\left(1-\tilde x_t^2\right)d_{20}^2(\tilde x_t)\left(2\tilde x_t'-1\right)Q_0(\tilde x_t)\notag\\
&-\frac{\tilde x_t}{\sqrt{6}\,\tilde \gamma}\bigg(\frac{16+25 \tilde \gamma+50 \tilde \delta}{10 }-\frac{(10+3 \tilde \gamma+6 \tilde \delta) \tilde x_t^2}{2 }+3 \tilde x_t^4\bigg)\bigg\},
\end{align}
where
\begin{align}
 \tilde\gamma&=\frac{8\qq'^2(s'-a)}{t(t-4\mpi^2)},\quad \tilde\delta=\frac{(t-2\mpi^2+2a)^2-4(s'-a)(2\qq'^2+2\mpi^2-s'-a)}{t(t-4\mpi^2)},\quad z_s'=\frac{z_t^2-\tilde\delta}{\tilde\gamma},\notag\\
\tilde x_t&=\frac{t+2s'-2\mpi^2}{\sqrt{t(t-4\mpi^2)}},\quad \tilde x_t'=\frac{\tilde x_t^2-\tilde\delta}{\tilde\gamma}=1+\frac{t}{2\qq'^2},
\end{align}
and
\begin{align}
 \Delta\tilde G_{JJ'}^{++\,(1)}(t,s')&=\frac{t}{8s'\qq'^2}(2J'+1)\delta_{J0}\Big(2\qq'^2\big(1-z_s'\big|_0\big)+t\Big)\bigg(\frac{2}{s'-\mpi^2}-\frac{1}{s'-a}\bigg)\frac{d_{11}^{J'}(z_s')}{1+z_s'}\bigg|_0,\notag\\
\Delta\tilde G_{JJ'}^{++\,(2)}(t,s')&=\frac{t^2}{8s'\qq'^2}(2J'+1)\delta_{J0}\Bigg\{(t-4\mpi^2)\notag\\
&\times\bigg\{\bigg(\frac{2}{s'-\mpi^2}-\frac{1}{s'-a}\bigg)\partial_t\frac{d_{11}^{J'}(z_s')}{1+z_s'}\bigg|_0-\frac{1}{(s'-\mpi^2)^2}\frac{d_{11}^{J'}(z_s')}{1+z_s'}\bigg|_0\bigg\}\notag\\
&-\bigg(\frac{2}{s'-\mpi^2}-\frac{1}{s'-a}\bigg)\partial_t(t'-4\mpi^2)\frac{d_{11}^{J'}(z_s')}{1+z_s'}\bigg|_0
+\frac{1}{(s'-\mpi^2)^2}(t'-4\mpi^2)\frac{d_{11}^{J'}(z_s')}{1+z_s'}\bigg|_0\Bigg\},\notag\\
 \Delta\tilde G_{JJ'}^{+-\,(1)}(t,s')&=-\frac{t}{2\qq'^2}(2J'+1)\delta_{J0}\bigg(\frac{2}{s'-\mpi^2}-\frac{1}{s'-a}\bigg)\frac{d_{1,-1}^{J'}(z_s')}{1-z_s'}\bigg|_0,\notag\\
\Delta\tilde G_{JJ'}^{+-\,(2)}(t,s')&=-\frac{t^2}{2\qq'^2}(2J'+1)\delta_{J0}\bigg\{\bigg(\frac{2}{s'-\mpi^2}-\frac{1}{s'-a}\bigg)\partial_t\frac{d_{1,-1}^{J'}(z_s')}{1-z_s'}\bigg|_0-\frac{1}{(s'-\mpi^2)^2}\frac{d_{1,-1}^{J'}(z_s')}{1-z_s'}\bigg|_0\bigg\},\notag\\
 \Delta\tilde G_{JJ'}^{-+\,(1)}(t,s')&=-\frac{t(t-4\mpi^2)}{2s'\qq'^2}(2J'+1)\frac{\delta_{J2}}{5\sqrt{6}}\bigg(\frac{2}{s'-\mpi^2}-\frac{1}{s'-a}\bigg)\frac{d_{11}^{J'}(z_s')}{1+z_s'}\bigg|_0,\notag\\
\Delta\tilde G_{JJ'}^{-+\,(2)}(t,s')&=-\frac{t^2(t-4\mpi^2)}{2s'\qq'^2}\frac{\delta_{J2}}{5\sqrt{6}}(2J'+1)\notag\\
&\times\bigg\{\bigg(\frac{2}{s'-\mpi^2}-\frac{1}{s'-a}\bigg)\partial_t\frac{d_{11}^{J'}(z_s')}{1+z_s'}\bigg|_0-\frac{1}{(s'-\mpi^2)^2}\frac{d_{11}^{J'}(z_s')}{1+z_s'}\bigg|_0\bigg\}.
\end{align}

\subsection{$\boldsymbol{t}$-channel}

The non-vanishing kernel functions for $J,J'\leq 2$ are
\begin{align}
 \tilde K_{00}^{++\,(0)}(t,t')&=\frac{t}{t'(t'-t)},\quad \tilde K_{00}^{++\,(1)}(t,t')=\frac{t^2}{t'^2(t'-t)}, 
\quad \tilde K_{00}^{++\,(2)}(t,t')=\frac{t^3}{t'^3(t'-t)},\notag\\
\tilde K_{02}^{++\,(0)}(t,t')&=\frac{5t(t+t'-4 \mpi^2+6 a)}{t'^2(t'-4\mpi^2)},\quad 
\tilde K_{02}^{++\,(1)}(t,t')=\frac{5t^2}{t'^2(t'-4\mpi^2)}, \quad  \tilde K_{02}^{++\,(2)}(t,t')=-\frac{10\mpi^2t^2}{t'^3(t'-4\mpi^2)},\notag\\
\tilde K_{22}^{++\,(0)}(t,t')&=\tilde K_{22}^{++\,(1)}(t,t')=\tilde K_{22}^{++\,(2)}(t,t')=\frac{t^2(t-4 \mpi^2)}{t'^2(t'-4\mpi^2)(t'-t)},\notag\\
\tilde K_{02}^{+-\,(0)}(t,t')&=\frac{5 \sqrt{6}\, t}{4 t' (t'-4 \mpi^2)},\quad 
\tilde K_{02}^{+-\,(1)}(t,t')=\frac{5 \sqrt{6}\, t^2}{4 t'^2 (t'-4 \mpi^2)},\quad
\tilde K_{02}^{+-\,(2)}(t,t')=\frac{5 \sqrt{6}\, t^3}{4 t'^3 (t'-4 \mpi^2)},\notag\\
\tilde K_{22}^{--\,(0)}(t,t')&=\frac{t(t-4 \mpi^2)}{t'(t'-4\mpi^2)(t'-t)},\quad \tilde K_{22}^{--\,(0)}(t,t')=\frac{t^2(t-4 \mpi^2)}{t'^2(t'-4\mpi^2)(t'-t)},\notag\\
\tilde K_{22}^{--\,(2)}(t,t')&=\frac{t^3(t-4 \mpi^2)}{t'^3(t'-4\mpi^2)(t'-t)}.
\end{align}

\setcounter{equation}{0}
\section{Omn\`es solutions for the $\boldsymbol{\gamma\gamma\to\pi\pi}$ partial waves}
\label{App:omnes_sol}

\subsection{$\boldsymbol{I=0}$}
\label{App:omnes_sol_I0}

The equations hold for $0<\delta(\tm)<\pi$, which all the $I=0$ partial waves satisfy. We only show the results for the once- and twice-subtracted version, as the reduction to the unsubtracted case is straightforward
\begin{align}
h_{0,+}(t)&=\tilde\Delta_{0,+}^{(1)}(t)+\frac{\mpi}{2\alpha}(\alpha_1-\beta_1)t\Omega_0(t)+\frac{t^{2}\Omega_0(t)}{\pi}\Bigg\{\int\limits_{4\mpi^2}^{\tm}\diff t'\frac{\sin\delta_0(t')\tilde\Delta_{0,+}^{(1)}(t')}{t'^2(t'-t)|\Omega_{0}(t')|}
+\int\limits_{\tm}^{\infty}\diff t'\frac{\Im h_{0,+}(t')}{t'^2(t'-t)|\Omega_{0}(t')|}\Bigg\},\notag\\ 
h_{0,+}(t)&=\tilde\Delta_{0,+}^{(2)}(t)+\frac{\mpi}{2\alpha}(\alpha_1-\beta_1)t(1-t\,\dot\Omega_0(0))\Omega_0(t)+\frac{\mpi}{24\alpha}(\alpha_2-\beta_2)t^2\Omega_0(t)\notag\\
&+\frac{t^{3}\Omega_0(t)}{\pi}\Bigg\{\int\limits_{4\mpi^2}^{\tm}\diff t'\frac{\sin\delta_0(t')\tilde\Delta_{0,+}^{(2)}(t')}{t'^3(t'-t)|\Omega_{0}(t')|}
+\int\limits_{\tm}^\infty\diff t'\frac{\Im h_{0,+}(t')}{t'^3(t'-t)|\Omega_{0}(t')|}\Bigg\},\notag\\ 
h_{2,+}(t)&=\tilde\Delta_{2,+}(t)
+\frac{t^{2}(t-4\mpi^2)\Omega_2(t)}{\pi}\Bigg\{\int\limits_{4\mpi^2}^{\tm}\diff t'\frac{\sin\delta_2(t')\tilde\Delta_{2,+}(t')}{t'^2(t'-4\mpi^2)(t'-t)|\Omega_{2}(t')|}
+\int\limits_{\tm}^\infty\diff t'\frac{\Im h_{2,+}(t')}{t'^2(t'-4\mpi^2)(t'-t)|\Omega_{2}(t')|}\Bigg\},\notag\\
h_{2,-}(t)&=\tilde\Delta_{2,-}^{(1)}(t)+\frac{t(t-4\mpi^2)}{10\sqrt{6}\mpi\alpha}(\alpha_1+\beta_1)\Omega_2(t)\notag\\
&+\frac{t^{2}(t-4\mpi^2)\Omega_2(t)}{\pi}\Bigg\{\int\limits_{4\mpi^2}^{\tm}\diff t'\frac{\sin\delta_2(t')\tilde\Delta_{2,-}^{(1)}(t')}{t'^2(t'-4\mpi^2)(t'-t)|\Omega_{2}(t')|}
+\int\limits_{\tm}^\infty\diff t'\frac{\Im h_{2,-}(t')}{t'^2(t'-4\mpi^2)(t'-t)|\Omega_{2}(t')|}\Bigg\},\notag\\
h_{2,-}(t)&=\tilde\Delta_{2,-}^{(2)}(t)+\frac{t(t-4\mpi^2)}{10\sqrt{6}\mpi\alpha}\Big[(\alpha_1+\beta_1)(1-t\,\dot\Omega_2(0))+\frac{t}{12}(\alpha_2+\beta_2)\Big]\Omega_2(t)\notag\\
&+\frac{t^{3}(t-4\mpi^2)\Omega_2(t)}{\pi}\Bigg\{\int\limits_{4\mpi^2}^{\tm}\diff t'\frac{\sin\delta_2(t')\tilde\Delta_{2,-}^{(2)}(t')}{t'^3(t'-4\mpi^2)(t'-t)|\Omega_{2}(t')|}
+\int\limits_{\tm}^\infty\diff t'\frac{\Im h_{2,-}(t')}{t'^3(t'-4\mpi^2)(t'-t)|\Omega_{2}(t')|}\Bigg\}.
\end{align}
The left-hand-cut functions $\tilde \Delta_{J,\pm}^{(i)}(t)$ are defined in \eqref{SDwave} and \eqref{Delta_tilde}.

\subsection{$\boldsymbol{I=2}$}
\label{App:omnes_sol_I2}

The solutions for $-\pi<\delta(\tm)<0$, which is true for all $I=2$ partial waves, read
\begin{align}
h_{0,+}(t)&=\tilde\Delta_{0,+}^{(1)}(t)+\frac{\mpi}{2\alpha}(\alpha_1-\beta_1)t\Omega_0(t)\frac{\tm-t}{\tm}\notag\\
&+\frac{t^{2}\Omega_0(t)(\tm-t)}{\pi}\Bigg\{\int\limits_{4\mpi^2}^{\tm}\diff t'\frac{\sin\delta_0(t')\tilde\Delta_{0,+}^{(1)}(t')}{t'^2(\tm-t')(t'-t)|\Omega_{0}(t')|}+\int\limits_{\tm}^{\infty}\diff t'\frac{\Im h_{0,+}(t')}{t'^2(\tm-t')(t'-t)|\Omega_{0}(t')|}\Bigg\},\notag\\ 
h_{0,+}(t)&=\tilde\Delta_{0,+}^{(2)}(t)+\frac{\mpi}{2\alpha}(\alpha_1-\beta_1)t\Big[1+\frac{t}{\tm}(1-\tm\,\dot \Omega_0(0))\Big]\Omega_0(t)\frac{\tm-t}{\tm}+\frac{\mpi}{24\alpha}(\alpha_2-\beta_2)t^2\Omega_0(t)\frac{\tm-t}{\tm}\notag\\
&+\frac{t^{3}\Omega_0(t)(\tm-t)}{\pi}\Bigg\{\int\limits_{4\mpi^2}^{\tm}\diff t'\frac{\sin\delta_0(t')\tilde\Delta_{0,+}^{(2)}(t')}{t'^3(\tm-t')(t'-t)|\Omega_{0}(t')|}
+\int\limits_{\tm}^\infty\diff t'\frac{\Im h_{0,+}(t')}{t'^3(\tm-t')(t'-t)|\Omega_{0}(t')|}\Bigg\},\notag\\ 
h_{2,+}(t)&=\tilde\Delta_{2,+}(t)+\frac{t^{2}(t-4\mpi^2)\Omega_2(t)(\tm-t)}{\pi}\Bigg\{\int\limits_{4\mpi^2}^{\tm}\diff t'\frac{\sin\delta_2(t')\tilde\Delta_{2,+}(t')}{t'^2(t'-4\mpi^2)(\tm-t')(t'-t)|\Omega_{2}(t')|}\notag\\
&+\int\limits_{\tm}^\infty\diff t'\frac{\Im h_{2,+}(t')}{t'^2(t'-4\mpi^2)(\tm-t')(t'-t)|\Omega_{2}(t')|}\Bigg\},\notag\\
h_{2,-}(t)&=\tilde\Delta_{2,-}^{(1)}(t)+\frac{t(t-4\mpi^2)}{10\sqrt{6}\mpi\alpha}(\alpha_1+\beta_1)\Omega_2(t)\frac{\tm-t}{\tm}\notag\\
&+\frac{t^{2}(t-4\mpi^2)\Omega_2(t)(\tm-t)}{\pi}\Bigg\{\int\limits_{4\mpi^2}^{\tm}\diff t'\frac{\sin\delta_2(t')\tilde\Delta_{2,-}^{(1)}(t')}{t'^2(t'-4\mpi^2)(\tm-t')(t'-t)|\Omega_{2}(t')|}\notag\\
&+\int\limits_{\tm}^\infty\diff t'\frac{\Im h_{2,-}(t')}{t'^2(t'-4\mpi^2)(\tm-t')(t'-t)|\Omega_{2}(t')|}\Bigg\},\notag\\
h_{2,-}(t)&=\tilde\Delta_{2,-}^{(2)}(t)+\frac{t(t-4\mpi^2)}{10\sqrt{6}\mpi\alpha}\Big\{(\alpha_1+\beta_1)\Big[1+\frac{t}{\tm}(1-\tm\,\dot \Omega_2(0))\Big]+\frac{t}{12}(\alpha_2+\beta_2)\Big\}\Omega_2(t)\frac{\tm-t}{\tm}\notag\\
&+\frac{t^{3}(t-4\mpi^2)\Omega_2(t)(\tm-t)}{\pi}\Bigg\{\int\limits_{4\mpi^2}^{\tm}\diff t'\frac{\sin\delta_2(t')\tilde\Delta_{2,-}^{(2)}(t')}{t'^3(t'-4\mpi^2)(\tm-t')(t'-t)|\Omega_{2}(t')|}\notag\\
&+\int\limits_{\tm}^\infty\diff t'\frac{\Im h_{2,-}(t')}{t'^3(t'-4\mpi^2)(\tm-t')(t'-t)|\Omega_{2}(t')|}\Bigg\}.
\end{align}
The sum rules discussed in Sect.~\ref{sec:I2_sumrule} are then
\begin{align}
 0&=\frac{\mpi}{2\alpha}(\alpha_1-\beta_1)\tm+\frac{\tm^2}{\pi}\Bigg\{\int\limits_{4\mpi^2}^{\tm}\diff t'\frac{\sin\delta_0(t')\tilde\Delta_{0,+}^{(1)}(t')}{t'^2(t'-\tm)|\Omega_{0}(t')|}+\int\limits_{\tm}^{\infty}\diff t'\frac{\Im h_{0,+}(t')}{t'^2(t'-\tm)|\Omega_{0}(t')|}\Bigg\},\notag\\
0&=\frac{\mpi}{2\alpha}(\alpha_1-\beta_1)\tm(1-\tm\,\dot \Omega_0(0))+\frac{\mpi}{24\alpha}(\alpha_2-\beta_2)\tm^2\notag\\
&+\frac{\tm^3}{\pi}\Bigg\{\int\limits_{4\mpi^2}^{\tm}\diff t'\frac{\sin\delta_0(t')\tilde\Delta_{0,+}^{(2)}(t')}{t'^3(t'-\tm)|\Omega_{0}(t')|}
+\int\limits_{\tm}^{\infty}\diff t'\frac{\Im h_{0,+}(t')}{t'^3(t'-\tm)|\Omega_{0}(t')|}\Bigg\},\notag\\
0&=\frac{\tm^{2}(\tm-4\mpi^2)}{\pi}\Bigg\{\int\limits_{4\mpi^2}^{\tm}\diff t'\frac{\sin\delta_2(t')\tilde\Delta_{2,+}(t')}{t'^2(t'-4\mpi^2)(t'-\tm)|\Omega_{2}(t')|}
+\int\limits_{\tm}^\infty\diff t'\frac{\Im h_{2,+}(t')}{t'^2(t'-4\mpi^2)(t'-\tm)|\Omega_{2}(t')|}\Bigg\},\notag\\
0&=\frac{\tm(\tm-4\mpi^2)}{10\sqrt{6}\mpi\alpha}(\alpha_1+\beta_1)\notag\\
&+\frac{\tm^{2}(\tm-4\mpi^2)}{\pi}\Bigg\{\int\limits_{4\mpi^2}^{\tm}\diff t'\frac{\sin\delta_2(t')\tilde\Delta_{2,-}^{(1)}(t')}{t'^2(t'-4\mpi^2)(t'-\tm)|\Omega_{2}(t')|}+
\int\limits_{\tm}^\infty\diff t'\frac{\Im h_{2,-}(t')}{t'^2(t'-4\mpi^2)(t'-\tm)|\Omega_{2}(t')|}\Bigg\},\notag\\
0&=\frac{\tm(\tm-4\mpi^2)}{10\sqrt{6}\mpi\alpha}\Big[(\alpha_1+\beta_1)(1-\tm\,\dot\Omega_2(0))+\frac{\tm}{12}(\alpha_2+\beta_2)\Big]\notag\\
&+\frac{\tm^{3}(\tm-4\mpi^2)}{\pi}\Bigg\{\int\limits_{4\mpi^2}^{\tm}\diff t'\frac{\sin\delta_2(t')\tilde\Delta_{2,-}^{(2)}(t')}{t'^3(t'-4\mpi^2)(t'-\tm)|\Omega_{2}(t')|}
+\int\limits_{\tm}^\infty\diff t'\frac{\Im h_{2,-}(t')}{t'^3(t'-4\mpi^2)(t'-\tm)|\Omega_{2}(t')|}\Bigg\}.
\end{align}

\end{strip}

\end{appendix}

\end{document}